\documentclass[prd, twocolumn,superscriptaddress, nofootinbib, notitlepage, floatfix]{revtex4-1}
\usepackage{xcolor}
\usepackage{graphicx}
\usepackage{wrapfig}
\usepackage{amsmath}
\usepackage{amsfonts}
\usepackage{amssymb}
\usepackage{multirow}
\usepackage{slashed}
\usepackage{physics}
\usepackage[colorlinks=true, linkcolor=blue, citecolor=blue, urlcolor=blue]{hyperref}

\DeclareRobustCommand{\Eq}[1]{Eq.~\eqref{eq:#1}}

\DeclareRobustCommand{\fig}[1]{Fig.~\ref{fig:#1}}

\DeclareRobustCommand{\app}[1]{App.~\ref{app:#1}}
\DeclareRobustCommand{\sec}[1]{Sec.~\ref{sec:#1}}

\DeclareRobustCommand{\tb}[1]{Table~\ref{tb:#1}}
\DeclareRobustCommand{\refcite}[1]{Ref.~\cite{#1}}

\newcommand\bet{\begin{table}}
\newcommand\eet[1]{\label{tb:#1}\end{table}}

\begin{document}
\widetext

\title{Pion form factor and charge radius from Lattice QCD at physical point}

\author{Xiang Gao}
\email{xgao@bnl.gov}
\affiliation{Physics Department, Tsinghua University, Beijing 100084, China}
\affiliation{Physics Department, Brookhaven National Laboratory, Upton, NY 11973, USA}
\author{Nikhil Karthik}
\affiliation{Department of Physics, College of William \& Mary, Williamsburg, VA 23185, USA}
\affiliation{Thomas Jefferson National Accelerator Facility, Newport News, VA 23606, USA}
\author{Swagato Mukherjee}
\affiliation{Physics Department, Brookhaven National Laboratory, Upton, NY 11973, USA}
\author{Peter Petreczky}
\affiliation{Physics Department, Brookhaven National Laboratory, Upton, NY 11973, USA}
\author{Sergey Syritsyn}
\affiliation{Department of Physics and Astronomy, Stony Brook University, Stony Brook, NY 11794, USA}
\affiliation{RIKEN-BNL Research Center, Brookhaven National Lab, Upton, NY, 11973, USA}
\author{Yong Zhao}
\affiliation{Physics Division, Argonne National Laboratory, Lemont, IL 60439, USA}
\affiliation{Physics Department, Brookhaven National Laboratory, Upton, NY 11973, USA}

\begin{abstract}
We present our results on the electromagnetic form factor of pion over a wide
range of $Q^2$ using lattice QCD simulations with Wilson-clover valence quarks and HISQ sea quarks. 
We study the form factor at the physical point with a lattice spacing $a=0.076$ fm. 
To study the lattice spacing and quark mass effects, we also present results for 300 MeV 
pion at two different lattice spacings $a=0.04$ and 0.06 fm. 
The lattice calculations at the physical quark mass appear to agree with the experimental
results. Through fits to the form factor, we estimate the charge radius of pion for physical pion mass
to be $\langle r_{\pi}^2 \rangle=0.42(2)~{\rm fm}^2$.
\end{abstract}
\date{\today}
\maketitle

\section{Introduction}\label{sec:intro}
Pion is one of the most prominent strongly-interacting particles next to the nucleon since it
is a Goldstone boson of QCD. For this reason, it is important to study the pion internal structure
and find out if there is a connection between its internal structure  and its Goldstone
boson nature. This issue is particularly relevant for understanding the origin of mass
generation in QCD, see e.g. discussions in Refs. \cite{Cui:2020dlm,Roberts:2020udq}.

Knowledge of internal structure of the pion is much more limited than that of the nucleon.
On the partonic level, the parton distribution function (PDF) of the pion has been studied through the global
analysis of the Drell-Yan production in pion-nucleon collisions and in tagged deep inelastic scattering (DIS),
for recent analyses see Refs. \cite{Barry:2018ort,Novikov:2020snp}. 
Recently, there have been many efforts in lattice QCD to study the pion PDF 
\cite{Chen:2018fwa,Sufian:2019bol,Joo:2019bzr,Sufian:2020vzb,Izubuchi:2019lyk,Gao:2020ito},
which have used the quasi-PDF in Large Momentum Effective Theory \cite{Ji:2013dva,Ji:2014gla}, 
the pseudo-PDF~\cite{Radyushkin:2017cyf,Orginos:2017kos} 
and current-current correlator~\cite{Braun:2007wv,Ma:2014jla, Ma:2017pxb} 
(also referred to as a ``good lattice cross-section'') approaches, 
see Refs. \cite{Cichy:2018mum,Zhao:2018fyu, Radyushkin:2019mye, Ji:2020ect} for recent reviews.
Lattice calculations of the lowest moments of pion PDF 
\cite{Best:1997qp,Guagnelli:2004ga,Capitani:2005jp,Abdel-Rehim:2015owa,Oehm:2018jvm,Alexandrou:2020gxs}
are also available and can be used as additional
constraints in the global analysis.

Form factor, defined as
\begin{equation}
\langle P_1 | J_{\mu} | P_2 \rangle = (P_1+P_2)_{\mu} F_{\pi}(Q^2),
\end{equation}
with $J_{\mu}$ being the electromagnetic current and $Q^2=-(P_2-P_1)^2$,
provide a different insight into pion structure, namely the 
charge distribution. It can be, in principle, measured in
electron-pion scattering. Generalized parton distribution (GPD) combine the information contained
in PDF and form factors and provide a three-dimensional image of a hadron. In the case of the nucleon,
the study of the GPDs is the subject of large experimental and theory efforts (see e.g. Ref. \cite{Dudek:2012vr} for
a recent review). Experimental study of the
pion GPD is far more challenging and will be only possible at Electron-Ion Collider (EIC), if at all.
Fortunately, GPDs can be studied on the lattice using LaMET, including pion GPDs 
\cite{Liu:2019urm,Chen:2019lcm,Lin:2020rxa,Alexandrou:2020zbe}.

Experimentally, the pion form factor
was measured by scattering of pions off atomic electrons in Fermilab \cite{Dally:1981ur,Dally:1982zk} and CERN \cite{Amendolia:1984nz,Amendolia:1986wj}.
This allowed determination of the pion form factor for momentum transfer $Q^2$ up to $0.253$ GeV$^2$ 
\cite{Dally:1981ur,Dally:1982zk,Amendolia:1984nz,Amendolia:1986wj}. 
For larger $Q^2$, one has to determine the pion form factor
from the electro-production of charged pions off nucleons. The corresponding experiments have been performed in Cornell~\cite{Bebek:1974ww,Bebek:1976qm,Bebek:1977pe}
DESY~\cite{Ackermann:1977rp,Brauel:1979zk}, and Jlab~\cite{Volmer:2000ek,Tadevosyan:2007yd,Horn:2006tm,Blok:2008jy,Huber:2008id}.
These determinations, however, were model-dependent. The recent determination of the pion form factor up to $Q^2$ of $2.45~{\rm GeV}^2$ is carried out by the $F_{\pi}$ collaboration using data both from DESY and JLab~\cite{Huber:2008id}. Experiments at the future EIC facility
will allow us to probe even higher $Q^2$ up to $30~{\rm GeV}^2$ and possibly see the partonic structure in an exclusive
elastic process and make contact with asymptotic large-$Q^2$ perturbative behavior \cite{Lepage:1979zb}. In the timelike region, the pion form factor can be determined by analyzing $e^{+} e^{-} \rightarrow \pi^{+} \pi^{-}$ process \cite{Colangelo:2018mtw} (see also references therein). This analysis also constrains the form factor in the spacelike region.

Lattice QCD calculations allow one to obtain the pion form factor from first principles, i.e.
without any model dependence, up to relatively large $Q^2$. Therefore, they will provide
an important cross-check for the experimental determinations. The first 
lattice calculations of the pion form factor date back to late 80s and were performed in the quenched approximation \cite{Martinelli:1987bh,Draper:1988bp}. More recently, lattice
calculations of the pion form factor have been performed with two flavors ($N_f=2$) of dynamical quarks
\cite{Brommel:2006ww,Frezzotti:2008dr,Aoki:2009qn,Brandt:2013dua,Alexandrou:2017blh},
with physical-mass strange- and two light-quark flavors ($N_f=2+1$) 
\cite{Bonnet:2004fr,Boyle:2008yd,Nguyen:2011ek,Fukaya:2014jka,Aoki:2015pba,Feng:2019geu,Wang:2020nbf}, 
as well as with a dynamical charm quark,
a strange quark and two flavors of the light quarks with nearly-physical masses ($N_f=2+1+1$) \cite{Koponen:2015tkr}.
Most of the lattice studies focused on the small $Q^2$ behavior of the pion form factor and the extraction
of the pion charge radius. The pion charge radius is very sensitive to the quark mass. Chiral perturbation theory
predicts a logarithmic divergence of the pion charge radius when the quark mass goes to zero \cite{Bijnens:1998fm}.
Therefore, one has to work at the physical quark mass or have calculations
performed in an appropriate range of quark masses to perform chiral extrapolations. Furthermore, studies have been performed for lattice spacing $a>0.09$ fm. Constrained by the analyticity and unitarity, the charge radius is correlated with the phase of form factors in the timelike region. It is proposed in \refcite{Colangelo:2020lcg} that high-precision determinations of the pion form factor and the charge radius have potential to shed light on the discrepancy of hadronic vacuum polarization (HVP) derived from $e^{+}+e^{-} \rightarrow$ hadron cross-sections and lattice calculations~\cite{Borsanyi:2020mff}.

The aim of this paper is to study the pion form factor in a wide range of $Q^2$. Therefore, we perform calculations
for small lattice spacings, namely $a=0.04$fm and $0.06$ fm, with valence pion mass of about $300$ MeV.
Furthermore, to study quark-mass effect, we also perform calculations at the physical pion mass, though at
somewhat larger lattice spacing, $a=0.076$ fm. Unlike previous studies, we also perform calculations for highly boosted
pion in order to extend them in the future to the pion GPD.

\section{Lattice setup}\label{sec:latset}
\begin{table*}
\centering
\begin{tabular}{|c|c|c|c|c|c|c|c|c|c|}
\hline
\hline
Ensemble:  &  $m_{\pi}^{val}$ (GeV) & $c_{sw}$ &$r_G$ fm &  $t_s/a$     & $n_z$ &  $n_i~(i=x,y)$ & $j_z$ & \#cfgs & (\#ex,\#sl) \cr
\hline
$a=0.076$ fm, $m_{\pi}^{sea}=0.14$ GeV, & 0.14 & 1.0372& 0.59     &6, 8, 10  & [0,3] & $\pm$1,$\pm$2       & 2     & 350 & $(5, 100)$ \cr
$64\times 64^3$                   &      &         &         &          &[4,7] & $\pm$1,$\pm$2       & 5     & 350 & $(5, 100)$ \cr
\cline{5-10}
                  &      &         &         &     20     &1 & $\pm$1,$\pm$2       & 2     & 350 & $(5, 100)$ \cr

\hline
$a=0.06$ fm, $m_{\pi}^{sea}=0.16$ GeV,  & 0.3  & 1.0336& 0.54    &8, 10, 12  & [0,1]  & $\pm$1,$\pm$2     & 0     & 100 & $(1, 32)$  \cr
$64\times 48^3$       &      &         &          &         & [2,3]  & $\pm$1,$\pm$2     & 2     & 525 & $(1, 32)$  \cr                   &      &         &          &         & [4,5]  & $\pm$1,$\pm$2     & 3     & 525 & $(1, 32)$  \cr
\hline
                                  &     &          &          &         & [0,1]  & $\pm$1            & 0     & 314 & $(3, 96)$  \cr
$a=0.04$ fm, $m_{\pi}^{sea}=0.16$ GeV   & 0.3 & 1.02868& 0.36     &9,12,     & [0,1]  & $\pm$2            & 0     & 314 & $(2, 64)$  \cr
$64\times 64^3$                   &     &          &     &    15,18     & [2,3]  & $\pm$1            & 2     & 564 & $(4,128)$  \cr
                                  &     &          &          &         & [2,3]  & $\pm$2            & 2    & 564 & $(3,96)$   \cr
\hline
\hline
\end{tabular}
\caption{The lattice parameters used in our calculations. Shown are the gauge ensembles used in our
study, the valence pion mass, the coefficient of the clover term, the size of the smeared Gaussian
sources, the source-sink separations, used in the analysis of the three-point functions,
the value of the momenta and with the corresponding boost parameters (see the main text).
The last two columns show the number of gauge configurations and the number of sources
in AMA (see the main text).
}
\label{tb:setup}
\end{table*}

In this study, we use Wilson-Clover action with hypercubic (HYP) \cite{Hasenfratz:2001hp} link smearing on (2+1)-flavor
$L_t \times L_s^3$ lattice ensembles generated 
by HotQCD collaboration \cite{Bazavov:2014pvz,Bazavov:2019www}
with highly-improved staggered quark (HISQ) sea action.
For the clover coefficient we use the tree-level tadpole improved 
value $c_{sw}=u_0^{-3/4}$, with $u_0$ being the HYP-smeared plaquette expectation value.
This setup is the same as the one used by us to study the valence parton distribution
of the pion \cite{Izubuchi:2019lyk,Gao:2020ito}. As in Refs. \cite{Izubuchi:2019lyk,Gao:2020ito},
we use two lattice spacings $a=0.04$ fm and $a=0.06$ fm and the valence pion mass of $300$ MeV.
The lightest pion mass for these gauge configurations is $m_{\pi}^{sea}=160$ MeV and
the lattice spacings were fixed with the $r_1$ scale \cite{Bazavov:2014pvz} using the value
$r_1=0.3106(18)$ fm \cite{Bazavov:2010hj}.
In addition, we performed calculations at a lattices spacings of $0.076$ fm and valence pion mass
of $140$ MeV using gauge configurations that correspond
to the lightest pion mass of $m_{\pi}^{sea}=140$ MeV \cite{Bazavov:2019www}. 
The lattice spacing was set by
the kaon decay constant, $f_K$ \cite{Bazavov:2019www}. The lattice ensembles used in this
study and the corresponding parameters are summarized in Table \ref{tb:setup}.
Due to the HISQ action, the taste splitting in the pion sector is small for 
lattice spacings $a\leq 0.076$ fm. For $a = 0.076$ the root mean square pion mass is only $15\%$ higher than the lightest
pion mass, while the heaviest pion mass is only $25\%$ above the lightest pion mass \cite{Bazavov:2019www}.
In what follows for $a=0.076$ fm ensemble, will will not make a difference between the sea and the valence
pion mass and refer to this ensemble as $m_{\pi}=140$ MeV ensemble or the ensemble with physical pion mass.
The effects of partial quenching will persist at finite lattice spacings but will go away in the continuum
limit.

To obtain the form factor we calculate the pion two-point and three-point functions.
We consider two-point functions defined as
\begin{equation}\label{eq:c2pt}
\begin{aligned}
&C_{\rm 2pt}^{ss'}(t;P_z)=\left\langle \pi_s(\mathbf{P},t) \pi_{s'}^\dagger(\mathbf{P},0) \right\rangle,
\end{aligned}
\end{equation}
where $\pi_s(\mathbf{P},t)$ are either smeared or point sources, $s=S,P$, with spatial
momentum 
$$\mathbf{P}=\frac{2 \pi}{a L_s} \cdot (n_x,n_y,n_z).$$
As in the previous studies \cite{Izubuchi:2019lyk,Gao:2020ito}, we used boosted Gaussian sources 
in Coulomb gauge with boost along the $z$-direction $k_z=2 \pi/(a L_s) \cdot (0,0,j_z)$. 
The radius of the Gaussian sources $r_G$ is also given in Table \ref{tb:setup}.
The three-point function is defined as 
\begin{equation}
C_{\rm 3pt}(\mathbf{P}^f,\mathbf{P}^i,\tau,t_s)=\left\langle \pi_S(\mathbf{P}^f,t_s) O_{\gamma_t}(\tau) \pi_S^\dagger(\mathbf{P}^i,0)\right\rangle,
\end{equation}
with
\begin{equation}
O_{\gamma_t}(\tau)=\sum_{\mathbf{x}} e^{-i (\mathbf{P}^f-\mathbf{P}^i) \mathbf{x} } 
\bigg{[}\overline{u}(x)\gamma_t u(x)-\overline{d}(x)\gamma_t d(x) \bigg{]},~
x=(\mathbf{x}, \tau)
\end{equation}
being the isovector component of the electric charge operator. Note that the isosinglet component
of the electric charge vanishes between the pion states. 
The initial momentum in the above expression is $\mathbf{P}^i= 2\pi/(a L_s)\cdot (0,0,n_z)$, while
the final momentum is $\mathbf{P}^f=\mathbf{P}=\mathbf{P}^i+\mathbf{q}$. 
The values of the momenta used in this study as well as the 
corresponding boost parameter $j_z$ are summarized in Table \ref{tb:setup}. 
We calculated the three-point functions for three values of the source-sink separations, $t_s$ for
the two coarser lattices. For the finest lattice we used four source-sink separations. The source-sink separations used in our study are also listed in Table \ref{tb:setup}.

The calculations of the two- and three-point functions were performed on GPUs 
with the QUDA multi-grid algorithm~\cite{Clark:2016rdz} used for the Wilson-Dirac operator inversions to get the quark propagators. 
We used multiple sources per configuration together with All Mode Averaging (AMA) technique ~\cite{Shintani:2014vja} 
to increase the statistics.
The  stopping criterion for AMA was set to be  $10^{-10}$ and $10^{-4}$ for the exact and sloppy inversions, respectively.
Since the signal is deteriorating with increasing momenta, we use different number of sources and number of
gauge configurations for different momenta. The number of gauge configurations and number of sources used in
the analysis are given in the last two columns of Table \ref{tb:setup} for each value of the momenta.

For the study of the form-factor, it is convenient to use the Breit frame, where $|\mathbf{P}^i|=|\mathbf{P}^f|$.
Using the Breit frame is essential when studying 
the GPD within LaMET \cite{Liu:2019urm,Chen:2019lcm,Lin:2020rxa,Alexandrou:2020zbe}, therefore we also calculated 
the pion form factor using the Breit frame. The parameters of this set-up are summarized in
Table \ref{tb:setup2}.
\begin{table*}
\centering
\begin{tabular}{|c|c|c|c|c|c|c|c|}
\hline
\hline
Ensemble &  $m_{\pi}^{val}$  & $t_s/a$  & $n_z^p$ & $n_i^p$ &  $n^q_i$ & \#cfgs & (\#ex,\#sl) \cr
$a,L_t\times L_s^3$ & (GeV) & & &$i=x,y$ & $i=x,y$ & &\cr
\hline
$a=0.06$ fm, $m_{\pi}^{sea}=0.16$,       & 0.3 &8,& 2 &$\pm$1 &$\mp$2 & 120 & $(1, 32)$\cr
$64\times 48^3$     &   &10& &&&&\cr  
\hline
$a=0.04$ fm, $m_{\pi}^{sea}=0.16$,      & 0.3 &9,12,& 2 &$\pm$1 &$\mp$2 & 120 & $(1, 32)$\cr
$64\times 64^3$     &   &15,18& &&&&\cr 
   
\hline
\hline
\end{tabular}
\caption{Two sets of measurements in the Breit frame on the two heavy-pion ensembles are shown. 
Using the notation similar to \tb{setup}, the initial pion state with transverse momentum 
$P_\bot^i=2\pi n^p_i/(L_s a)$, 
has the same energy as the final state with momentum 
$\mathbf{P}^f$ = $\mathbf{P}^i$ + $\mathbf{q}$.
}
\label{tb:setup2}
\end{table*}

\section{Two-point function analysis}\label{sec:c2pt}
\begin{figure*}
\centering
\includegraphics[width=0.4\textwidth]{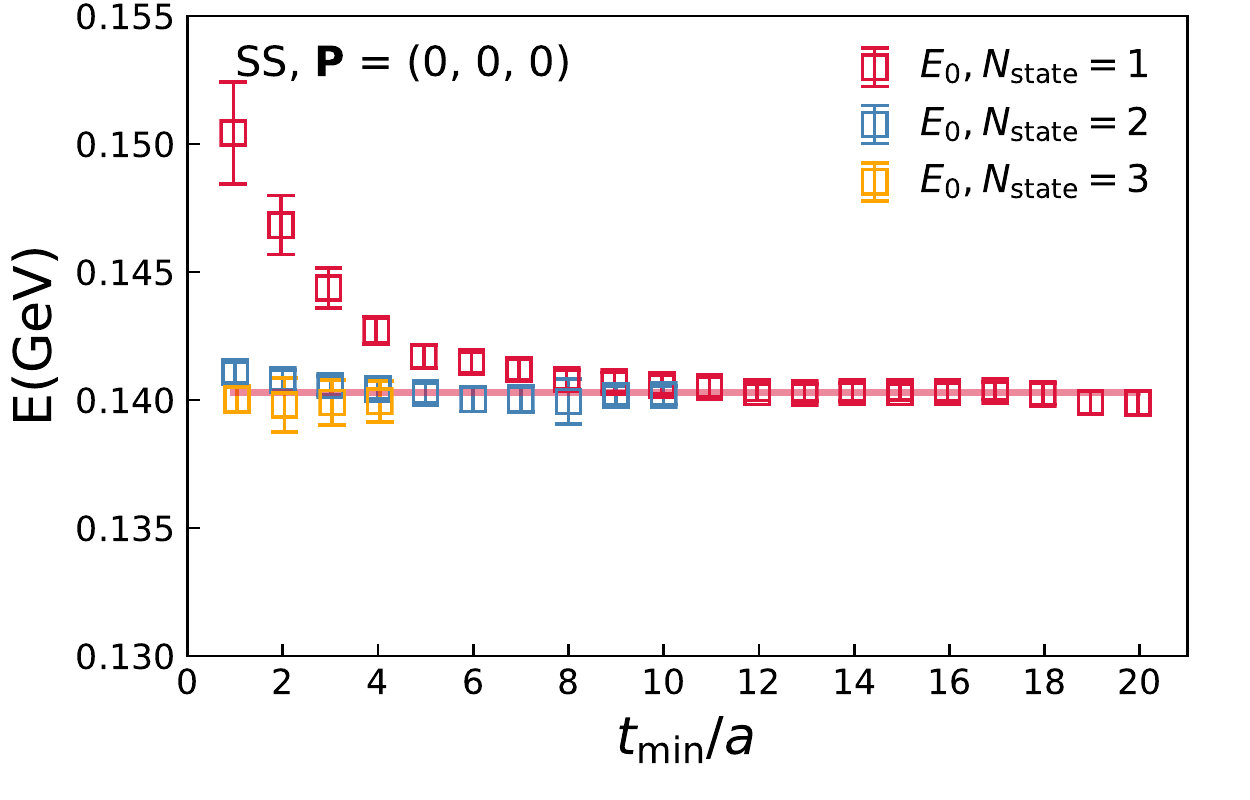}
\includegraphics[width=0.4\textwidth]{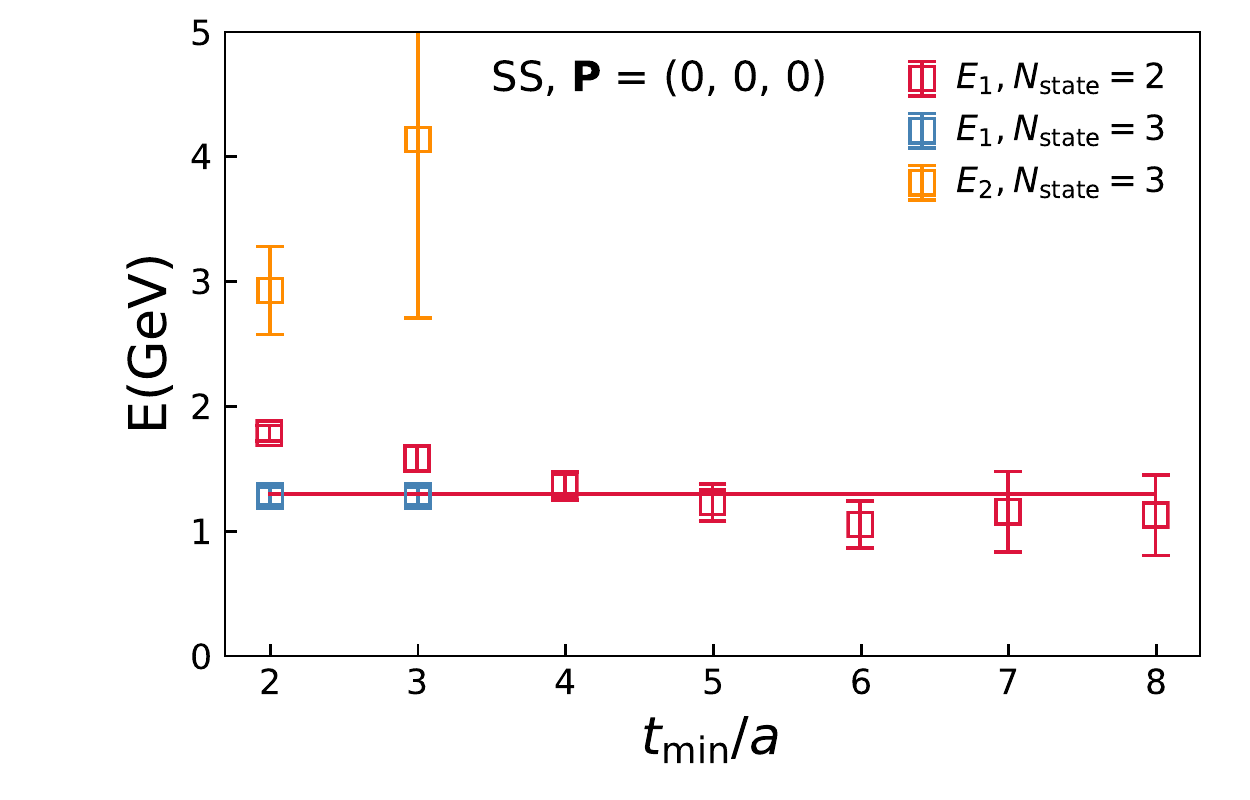}
\includegraphics[width=0.4\textwidth]{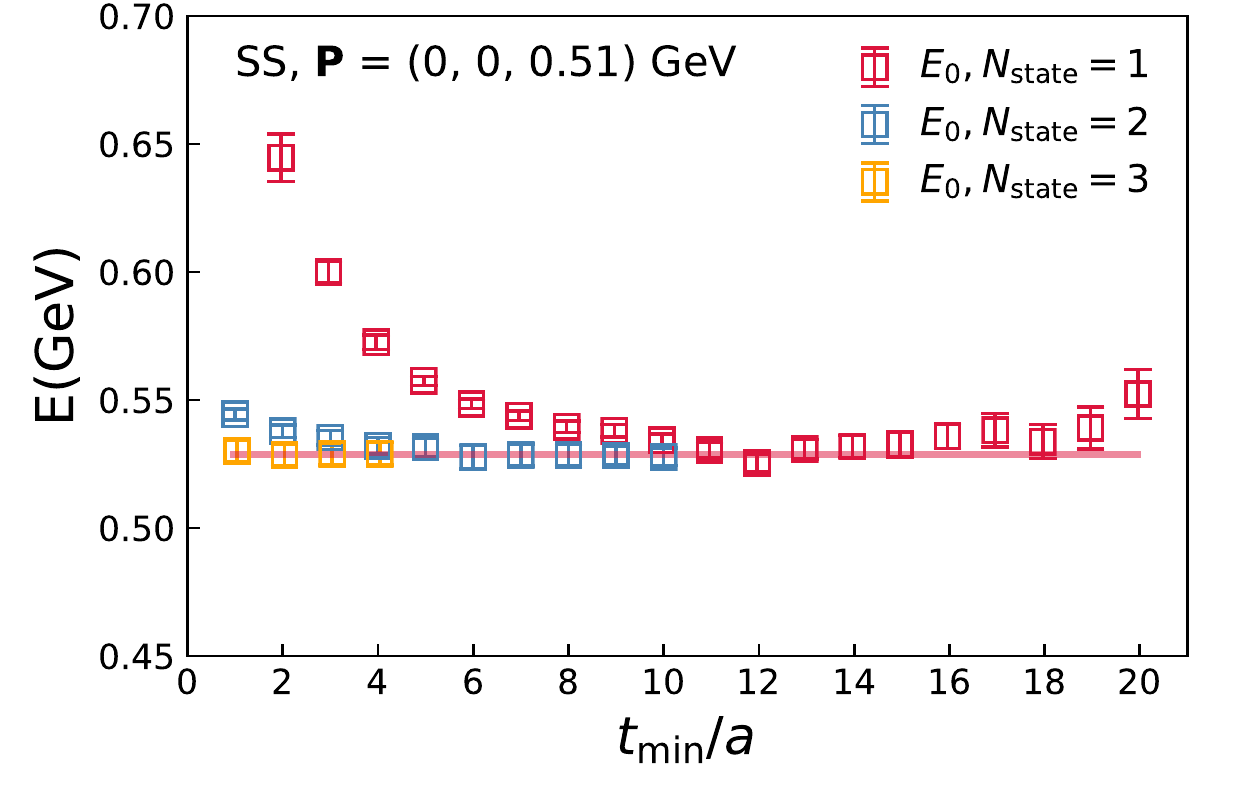}
\includegraphics[width=0.4\textwidth]{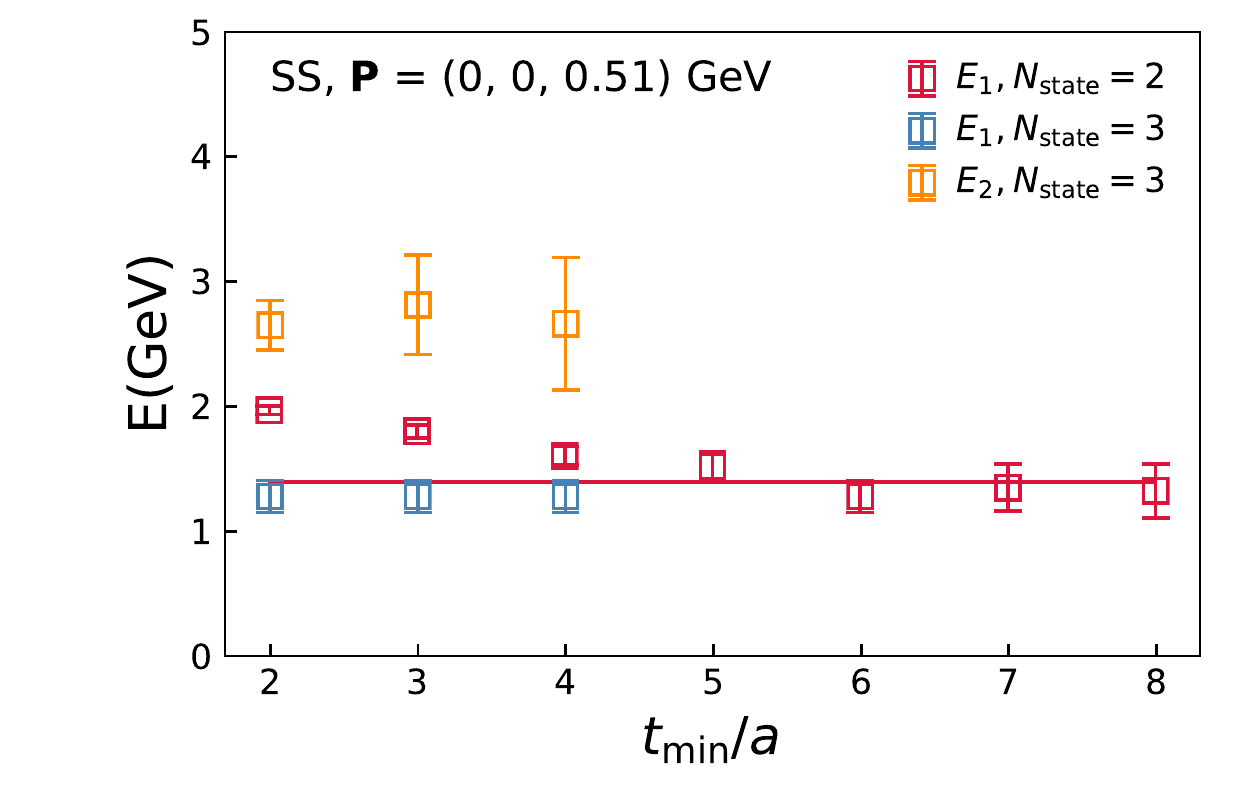}
\includegraphics[width=0.4\textwidth]{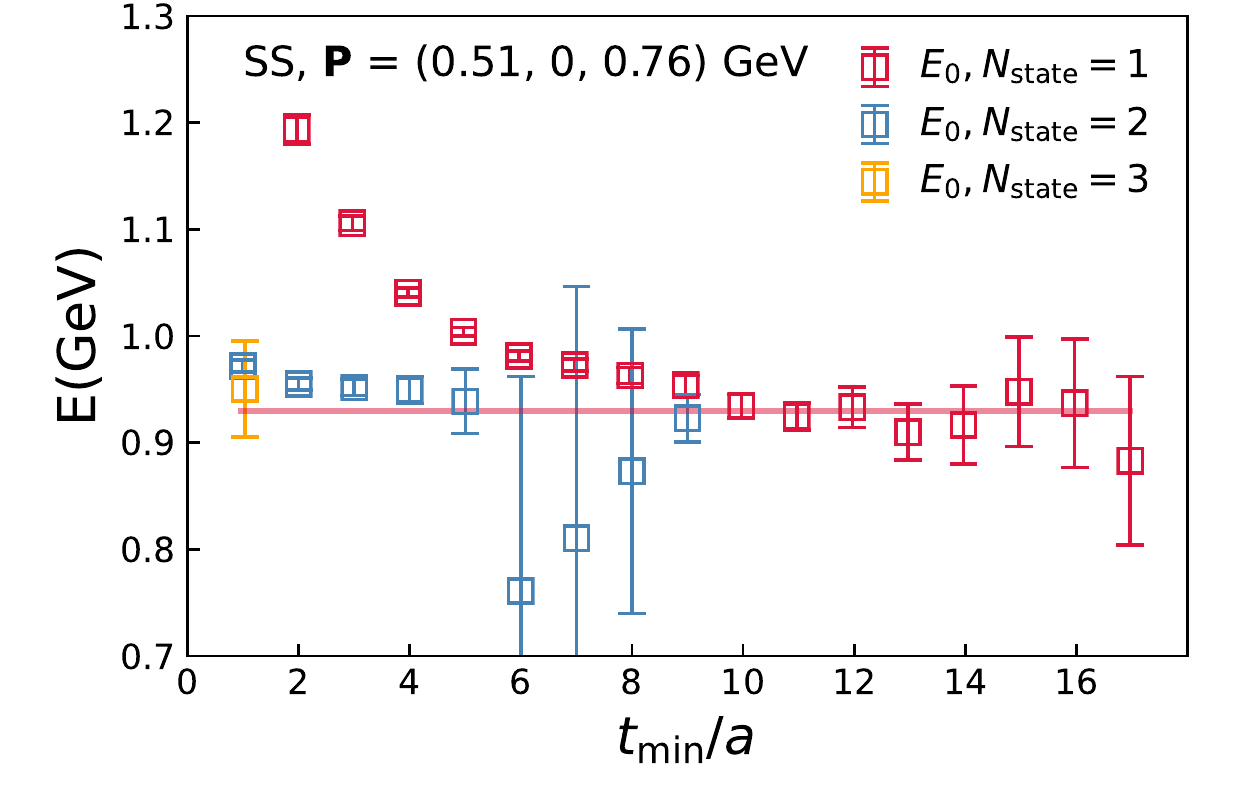}
\includegraphics[width=0.4\textwidth]{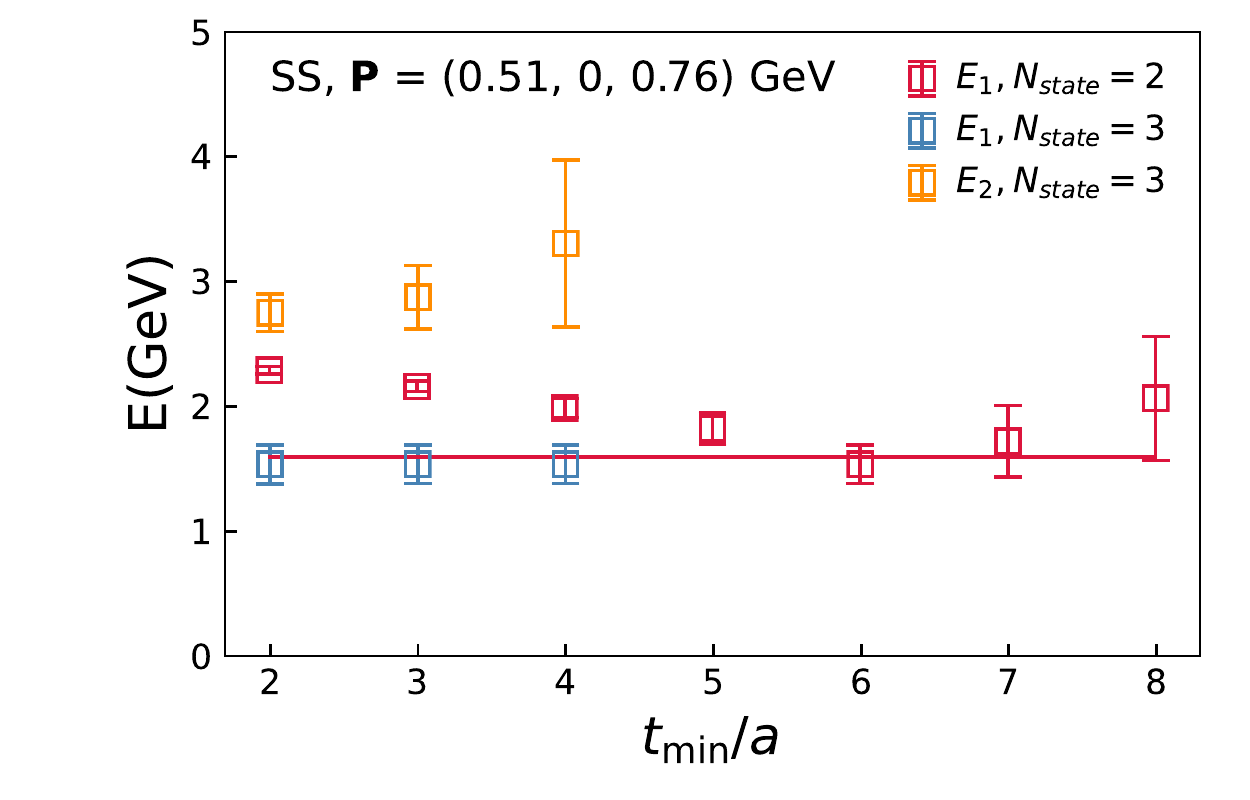}
\caption{$E_0$ from N-state fits (left) and $E_1$, $E_2$ from constrained 2-state and 3-state fits (right) for three different momenta are shown as functions of $t_{\rm{min}}$. The lines are computed from the dispersion relation $E(\mathbf{P})=\sqrt{\mathbf{P}^2+E(\mathbf{P}=0)^2}$, with $E(\mathbf{P}=0)$ to be 0.14 GeV for $E_0$ and 1.3 GeV for $E_1$. As can be observed, the $E_0$ and $E_1$ reach a plateau for large enough $t_{\rm min}$.\label{fig:c2ptfit}}
\end{figure*}
\begin{figure*}
\includegraphics[width=0.35\textwidth]{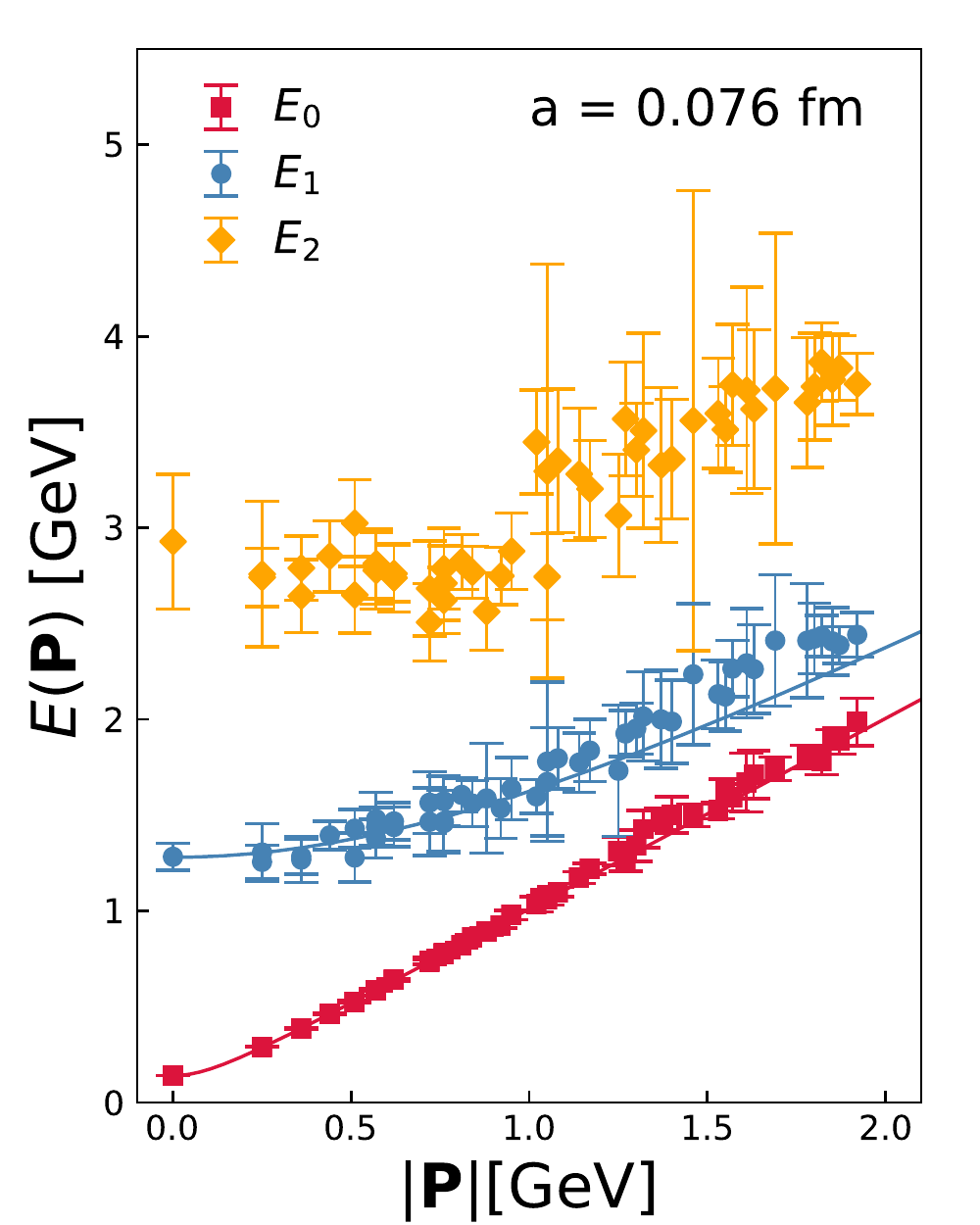}
\includegraphics[width=0.35\textwidth]{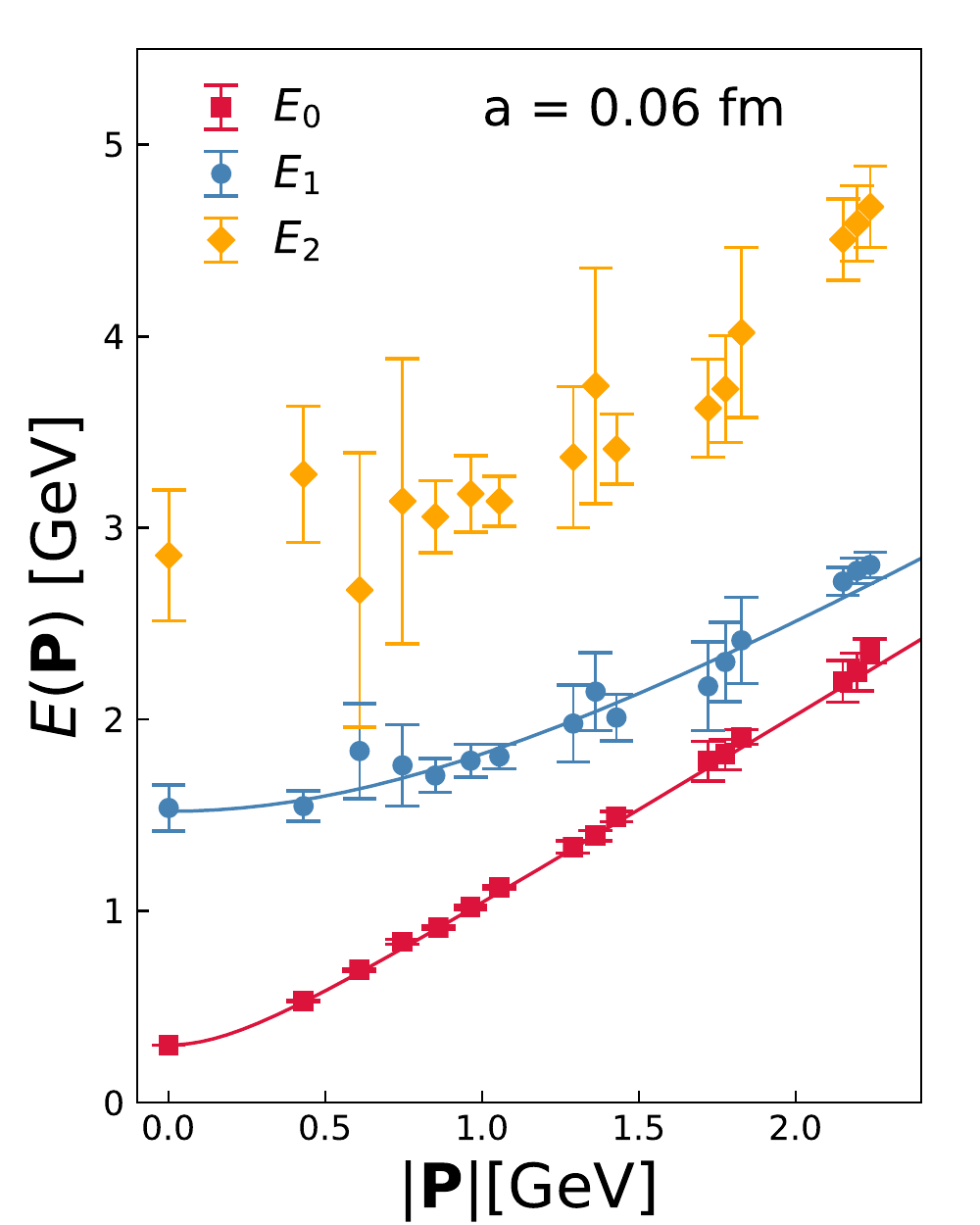}
\caption{Dispersion relation determined by the plateau of \fig{c2ptfit} for the physical pion mass ensemble (left) and a = 0.06 fm ensemble (right). The lines are dispersion relation calculated by $E(\mathbf{P})=\sqrt{\mathbf{P}^2+E(\mathbf{P}=0)^2}$. \label{fig:disp}}
\end{figure*}

Since the source-sink separation values used in this study are not very large, it is important to quantify the contributions of the excited states when extracting pion matrix elements. This in turn requires a detailed study of the pion two-point functions. For $a=0.04$ fm and $0.06$ fm lattices and $m_{\pi}^{val}=300$ MeV, the pion two-point functions have been studied for different momenta along the $z$-direction in Refs. \cite{Izubuchi:2019lyk,Gao:2020ito}. Furthermore, this analysis was very recently extended to include momenta also along the $x$ and $y$-directions  for $a=0.04$ fm \cite{Gao:2021hvs}. We have extended this analysis to $a=0.076$ fm and the physical pion mass.

The pion two-point function in  \Eq{c2pt} has the following spectral decomposition:
\begin{equation}\label{eq:spectr}
    C_{\rm 2pt}^{ss'}(t)=\sum_{n=0}^{N_{\rm state}-1} A_n^s A_n^{s'*} (e^{-E_n t}+e^{-E_n (aL_t-t)}),
\end{equation}
where $E_{n+1}\textgreater E_{n}$, with $E_0$ being the energy of the pion ground state. $A_n$ is the overlap factor $\langle \Omega|\pi_s|n\rangle$ of the state $n$ and the state created by operator $\pi_s$ from the vacuum state $|\Omega\rangle$. Thanks to the Gaussian smearing, the excited state contribution is suppressed. So we truncate the \Eq{spectr} up to $N_{\rm state} = 3$ and then fit the data in a range of $t\in [t_{\rm min},\, a L_t/2]$. 
    In the left panels of \fig{c2ptfit}, we show the extracted $E_0$ for three different momenta. As one can see, the ground-state energies, $E_0$ reach a plateau when $t_{\rm min}\gtrsim 10a$, $5a$ and $2a$ for 1-state, 2-state and 3-state fits, respectively.
The horizontal lines in the plots are computed from the dispersion relation $E_0(\mathbf{P})=\sqrt{\mathbf{P}^2+m_\pi^2}$. 
Here the value of $m_\pi$ was obtained by considering the pion masses from the fits with $t_{\rm min}\in [10a, 20a]$, and then fitting these results to a constant. The fit to a constant has $\chi^2_{d.o.f} = 0.2$, i.e. there is
no statistically significant $t_{min}$ dependence of the pion mass. The ground-state energies for different momenta agree with the horizontal lines for sufficiently large $t_{min}$, i.e. follow the dispersion relation.
Thus for the determination of the next energy level, we can fix the ground-state energy $E_0$ to be from the dispersion relation, and perform a 3-state fit. Interestingly, as shown in right panels of \fig{c2ptfit}, we can also observe plateaus for $E_1$ when $t_{\rm min}\textgreater$5a. The energy of the first excited state also follows the dispersion relation $E_1(\mathbf{P})=\sqrt{\mathbf{P}^2+m_{\pi'}^2}$ with $m_{\pi'}$ = 1.3 GeV. This could imply that the first excited state is  single particle state, namely the first radial excitation of the pion $\pi$(1300) \cite{Gao:2021hvs}. 
We cannot rule out, however, the possibility that
it is a multi-pion states within the large errors. 
Since the first excited state energy, $E_1$ does not reach a plateau for $t_{\rm min}\textless$5a, we conclude that for $t/a< 5$ the contribution of higher excited states in the two-point function is significant. Therefore, we need to consider 3-state fits for these $t$ values. To perform a 3-state fit, we fix $E_0$ to the dispersion relation and put a prior to $E_1$ using the best estimates from SS and smeared-point (SP) correlators \cite{Gao:2020ito} together with the errors from the 2-state fit. This way we get the third excited state energy, $E_2$, which does not depend on $t_{min}$ within the statistical errors. However, the value of $E_2$ is very large, about 3 GeV. This implies that $E_2$ does not actually refer to a single state but rather to a tower of many higher excited states. The situation is similar for other two 300 MeV ensembles \cite{Gao:2020ito}.

Now we understand that a 2-state spectral model can describe our two-point functions well when $t_{\rm min}\gtrsim 5a$, while 3-state can describe $t_{\rm min}\gtrsim 2a$. This will be important to keep in mind when analyzing the three-point function and pion matrix elements in the next section. To summarize this section, in \fig{disp} we show the dispersion relation obtained from the above analysis. We also extended the analysis for $a=0.06$ fm \cite{Gao:2020ito} by including additional momenta with non-zero components along the $x$ and $y$-directions. The corresponding results are also shown in \fig{disp}. We clearly see the effect of the quark masses. For the larger quark mass ($a=0.06$ fm) the excited state is about 200 MeV higher than the physical point ($a=0.076$ fm). This fact again suggests that
the first excited state is the radial excitation
of the pion. One of the reason we do not have multi-pion states entering the two-point correlation function is the use of Gaussian
sources. These sources have poor overlap with the scattering states. 


\section{Extraction of bare matrix elements of pion ground state}\label{sec:c3pt}
\begin{figure*}
\includegraphics[width=0.4\textwidth]{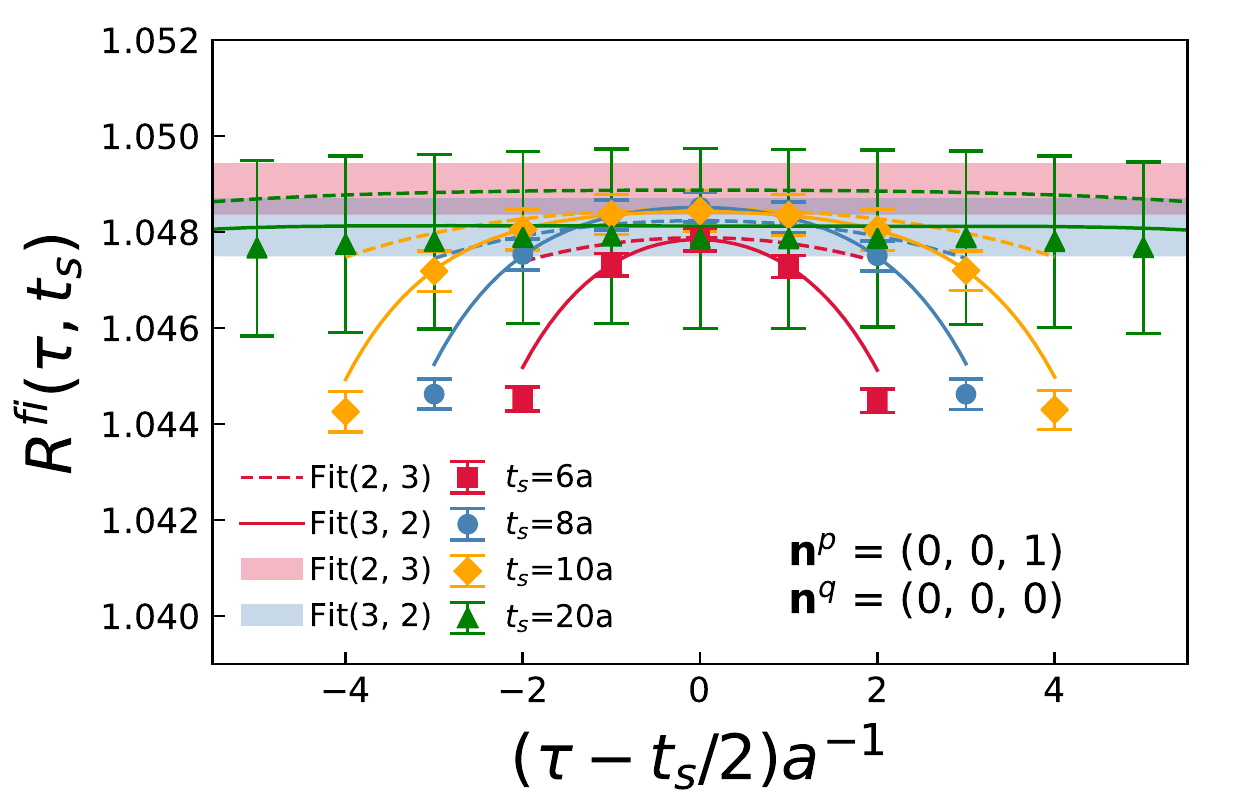}
\includegraphics[width=0.4\textwidth]{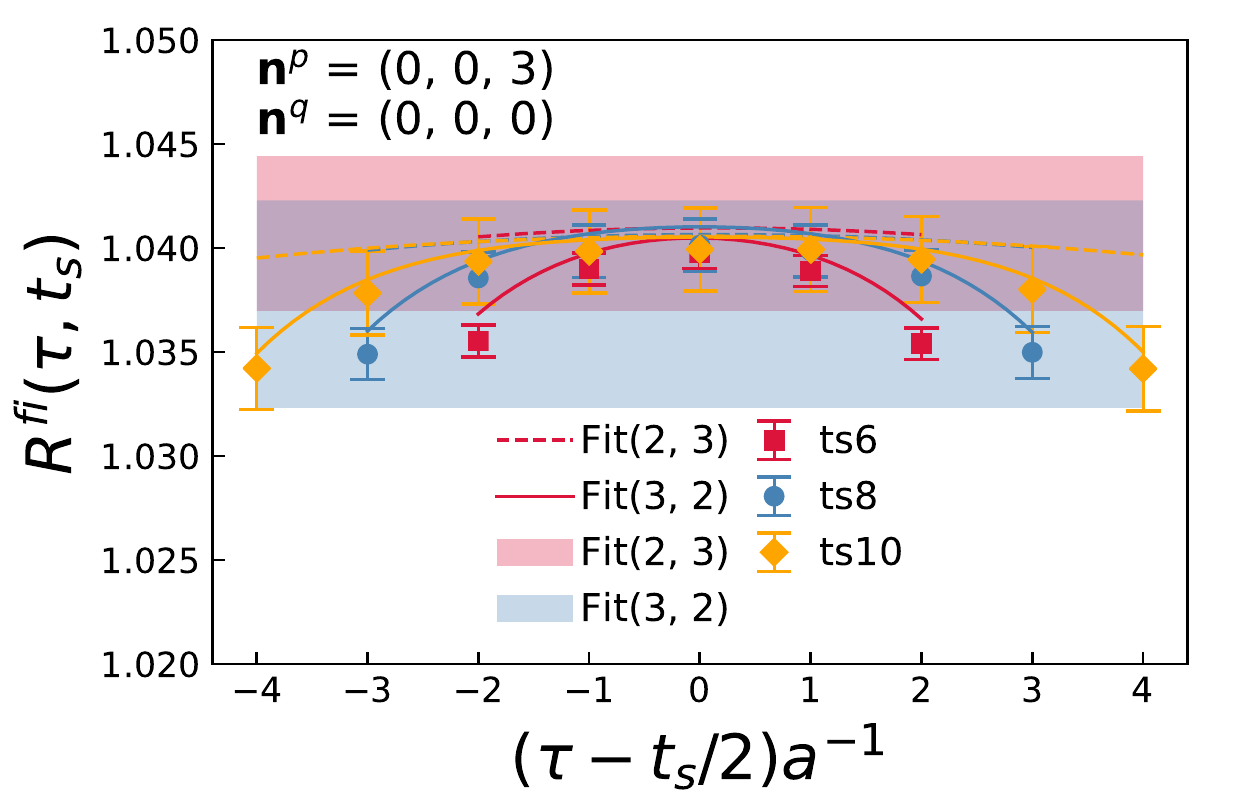}
\includegraphics[width=0.4\textwidth]{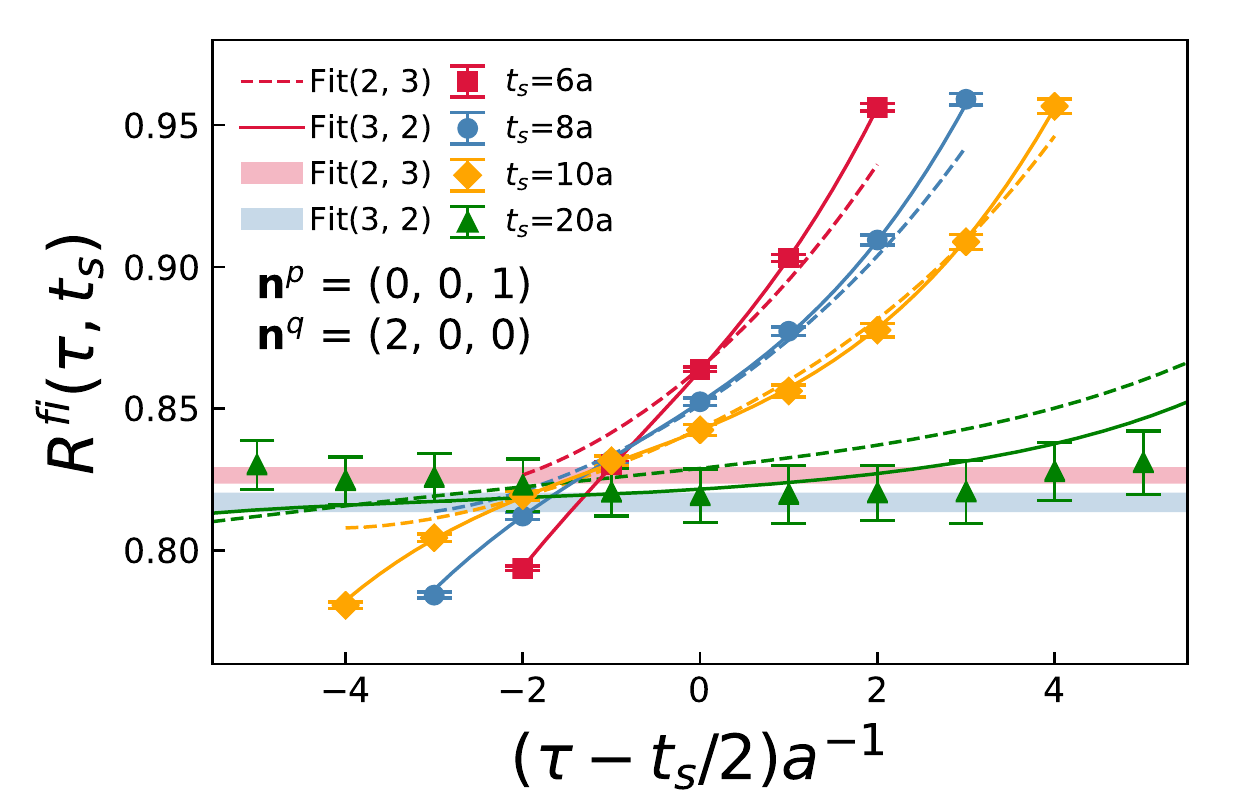}
\includegraphics[width=0.4\textwidth]{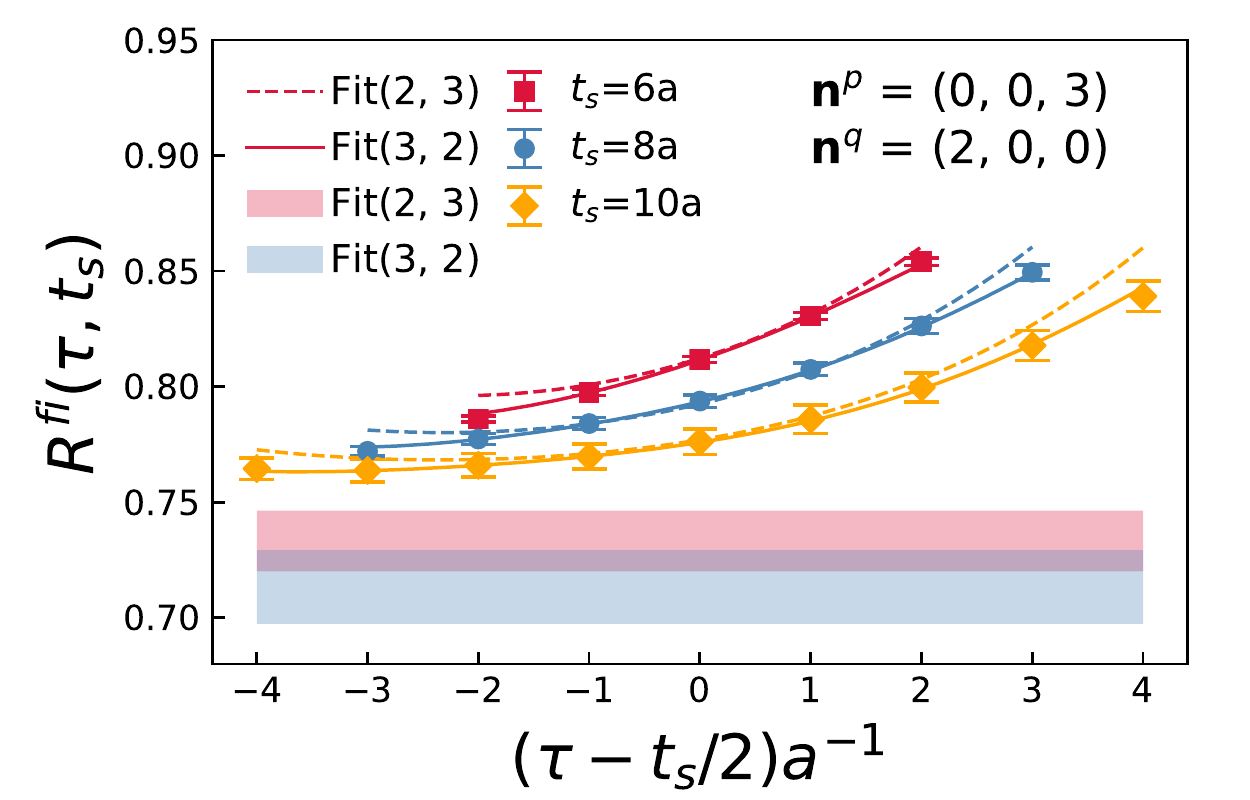}
\includegraphics[width=0.4\textwidth]{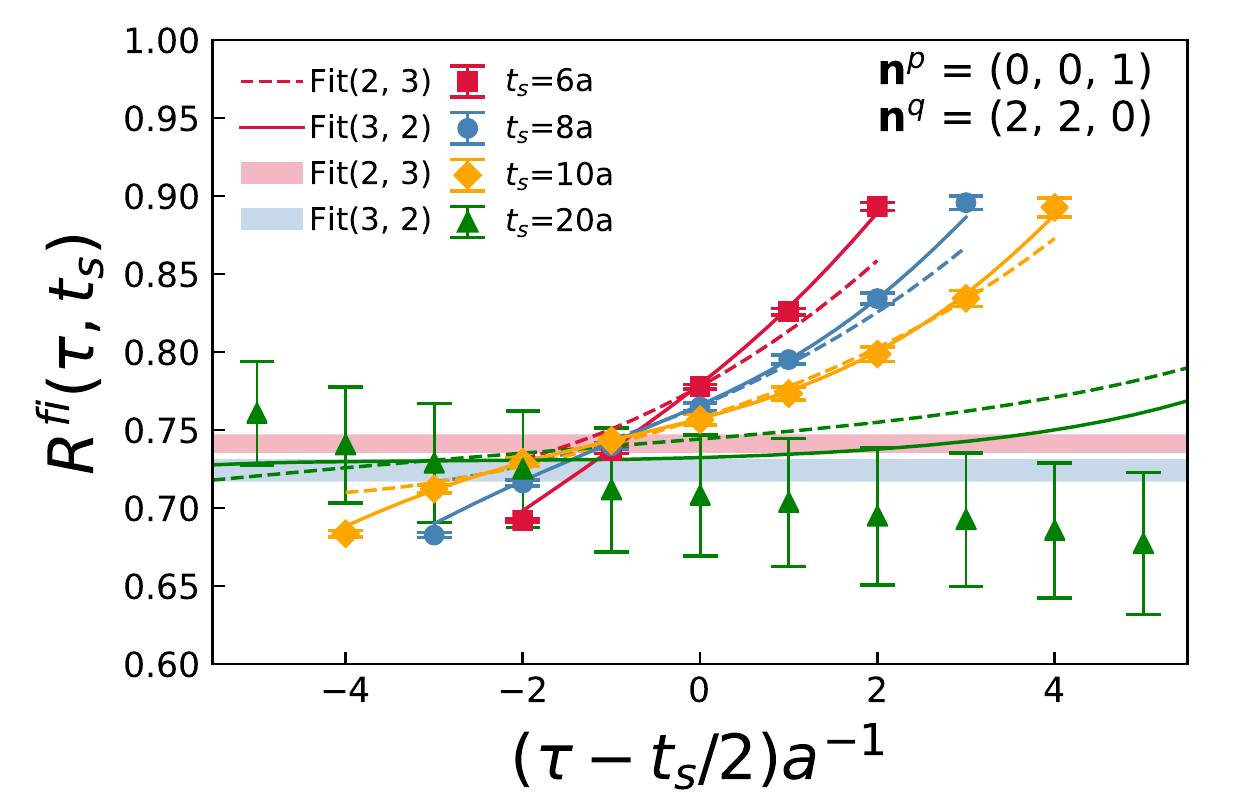}
\includegraphics[width=0.4\textwidth]{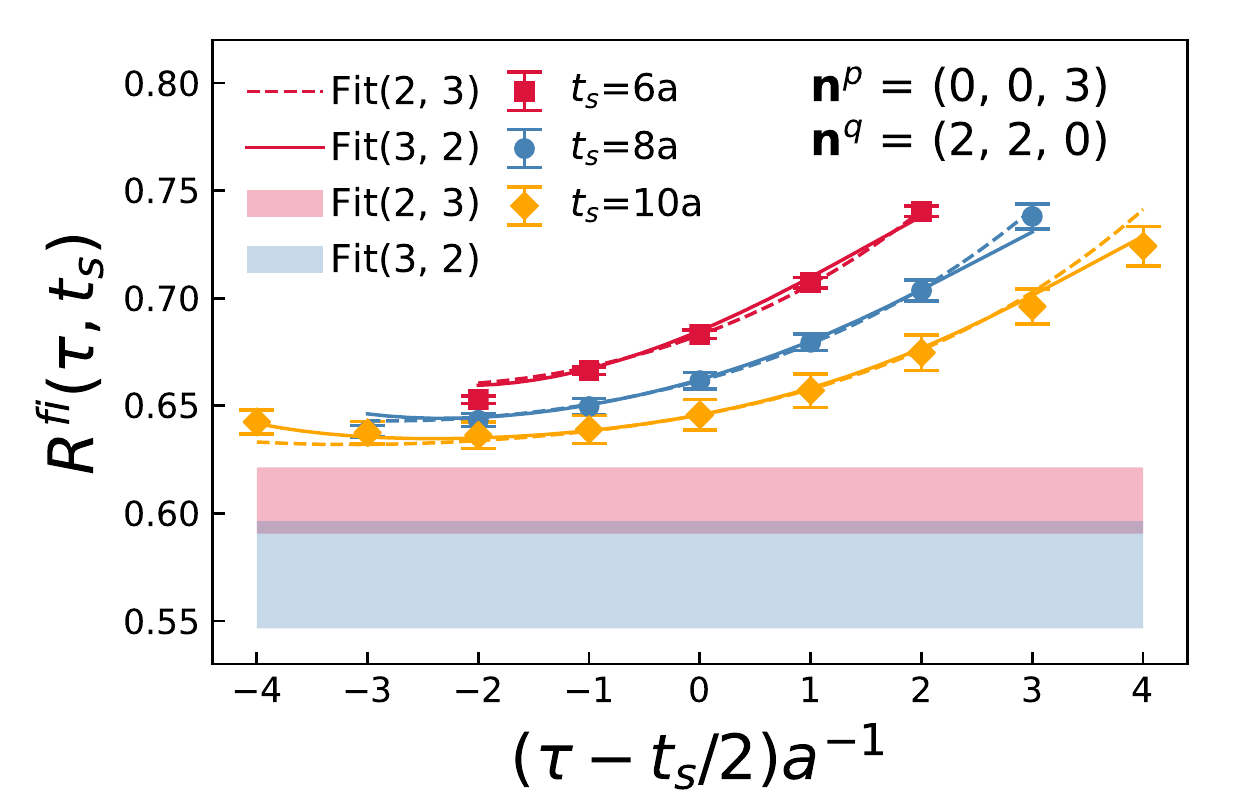}
\caption{$R^{fi}(\tau,t_s)$ for $\mathbf{n^p}_i = (0,0,1)$ (left) and $(0,0,3)$ (right) for $\mathbf{n^q}=(0,0,0),~(2,0,0),~(2,2,0)$ of physical ensemble are shown. The curves are reconstructed from the central value of multi-state fit Fit(2,3) (dashed) and Fit(3,2) (solid), and the bands are the estimated bare matrix elements from bootstrap method. \label{fig:ratiofit}}
\end{figure*}

To obtain the bare pion form factor we consider the following standard ratio of the three-point and two-point
pion correlation functions \cite{Capitani:1998ff,Wilcox:1991cq}
\begin{equation}\label{eq:ratio}
\begin{aligned}
   &R^{fi}(\tau,t_s) \equiv
   \frac{2\sqrt{P^f_0P^i_0}}{P^f_0+P^i_0} \frac{C_{\rm 3pt}(\mathbf{P}^f,\mathbf{P}^i,\tau,t_s)}{C_{\rm 2pt}(t_s,\mathbf{P}^i)}\\
   &\times \left(\frac{C_{\rm 2pt}(t_s-\tau,\mathbf{P}^f)C_{\rm 2pt}(\tau,\mathbf{P}^i)C_{\rm 2pt}(t_s,\mathbf{P}^i)}{C_{\rm 2pt}(t_s-\tau,\mathbf{P}^i)C_{\rm 2pt}(\tau,\mathbf{P}^f)C_{\rm 2pt}(t_s,\mathbf{P}^f)} \right)^{1/2}.
\end{aligned}
\end{equation}
This ratio gives the bare  pion form factor in the limit $\{\tau,\,(t_s-\tau)\} \rightarrow \infty$:
$h_B(P_f,P_i)=\lim_{\{\tau,\,(t_s-\tau)\}\rightarrow \infty} R^{fi}(\tau,t_s)$.

As explained in \sec{latset}, we calculated the three-point functions with $\mathbf{P}^i$ along 
the $\hat z$ direction, and multiple values of momentum transfer $\mathbf{q}=\mathbf{P}^f-\mathbf{P}^i$ for each $\mathbf{P}^i$. 
Thus there is no difference for $\mathbf{q}$ with same magnitude of the transverse momentum transfer. 
In other words, there should be transverse symmetry for the three-point function data. 
We find that indeed our numerical results for $R^{fi}(\tau,t_s)$ with same $|n^q_x|$ and $|n^q_y|$ 
are consistent within the error. Therefore, we average the three-point functions data 
with same magnitude of the transverse momentum transfer in the following analysis.

Since the temporal extent of our lattices is not large, it is important to consider thermal state contaminations, also called wrap-around effects
caused by the periodic boundary condition in time~\cite{Gao:2020ito}.
To remove the wrap-around effects in the two-point function 
we replaced $C_{\rm 2pt}(t)$ by $C_{\rm 2pt}(t)-A_0e^{-E_0(aL_t-t)}$ using the best estimate of $A_0$ and $E_0$
from the two-point function analysis. 
To understand wrap-around effects in the three-point function we consider the spectral decomposition of 
$C_{\rm 3pt}$ in \Eq{ratio} 
 \begin{equation}\label{eq:3pt_g}
 \begin{aligned}
&\langle \pi_S(\mathbf{P}^f,t_s) O_{\gamma_t}(\tau) \pi_S^\dagger(\mathbf{P}^i,0)\rangle \\
&= \sum_{m,n,k}\langle m|\pi_S|n\rangle\langle n|O_{\gamma_t}|k\rangle\langle k|\pi_S^\dagger|m\rangle \times\\
&
e^{-\tau E_{k}}e^{-(t_s-\tau)E_n}e^{-(aL_t-t_s) E_m},
 \end{aligned}
\end{equation}
where $m,n,k = \Omega,~0,~1,~\dots $, with 0 being the pion ground state. 
In general, terms with non-zero $E_m$ will be highly suppressed by $e^{-(a L_t-t_s) E_m}$ (we assume $E_{\Omega}=0$).
Therefore, in most studies such terms are neglected.
However for the $P=0$ case $e^{-(a L_t-t_s) E_m(P=0)}=e^{-aL_tm_\pi}$ is not very small.
We have $e^{-aL_tm_\pi}\sim$ 0.03, 0.003, 0.02 
for a = 0.076, 0.06 and 0.04 fm lattices, respectively. 
On the other hand,
for non-zero momenta the  terms proportional to $e^{-(a L_t-t_s) E_m}$ are smaller than 0.003 and can be neglected.
Therefore, for $a=0.04$ fm and $0.076$ fm calculations we only consider non-zero momenta and
limit the sum over index $m$ in Eq. (\ref{eq:3pt_g}) to include only the vacuum
state in what follows.
We need, however, to consider the wrap-around effects when dealing with the renormalization, as discussed in
the next section. 

In this work, we use multi-state fit to extract the bare matrix elements 
of the ground state $\langle  P^f|O_{\gamma_t}| P^i \rangle \equiv \langle 0 P^f|O_{\gamma_t}| P^i 0\rangle$
by inserting the spectral decomposition of the two-point function in Eq. (\ref{eq:spectr})
and the three-point function in Eq. (\ref{eq:3pt_g}) with $m=\Omega$, and the sum over $n$ truncated to $N_{state}$ terms. Furthermore, 
we take the best estimate of $A_n$ and $E_n$ from the two-point function analysis.
and put them into \Eq{ratio}. 
In the following, we will refer to this method as $\textup{Fit}(N_{state},n_{sk})$, in which $N_{state}$ is the number of states
in the corresponding two-point function analysis
and $n_{sk}$ labels how many $\tau$ points are skipped on the two sides of $t_s$. 
We consider $N_{state}=2$ and $N_{state}=3$ that have four and nine fit parameters, respectively.

We perform multi-state fit using bootstrap method with time separations $t_s$ = 6a, 8a, 10a.
The data with $t_s$ = 20a and $\mathbf{n^p}_i = (0,0,1)$ are 
used only to cross-check our analysis. 
Since the ratio defined in \Eq{ratio} is a derived quantity not defined
on a single gauge configuration we used uncorrelated fits. 
The statistical correlation between
the different data points is taken into account through the bootstrap procedure.
In \fig{ratiofit}, we show the examples of ratio $R^{fi}(\tau,t_s)$ as well as the 2-state and 3-state fit results. As one can see, for large momentum with large statistical errors, the reconstructed curves go through the data points well, and the 2-state and 3-state fit results are consistent with each other. However, this is not the case for smaller momentum, where the data are more precise. The 3-state fit is required to describe the ratio data with $\chi^2/dof <$ 1, while the 2-state fit result in $\chi^2/dof \gg$ 1. Thus for the following analysis, we will take the 3-state fit results as the central value and use the corresponding statistical errors. However, even when using the 3-state fit there is no guarantee that we are free from excited state contamination. Therefore, we take the difference between the 2-state fit and the 3-state fit results as the systematic errors in the following analysis. It can be also observed that the data points of $t_s$ = 20a show plateau around $t_s/2$ within the errors, and are also consistent with the 3-state fit results, which support our estimate of bare matrix elements. In \app{app2}, we discuss the plateau fit results using $t_s$ = 20a data.

\section{The pion form factors}
\begin{figure}
\includegraphics[width=0.45\textwidth]{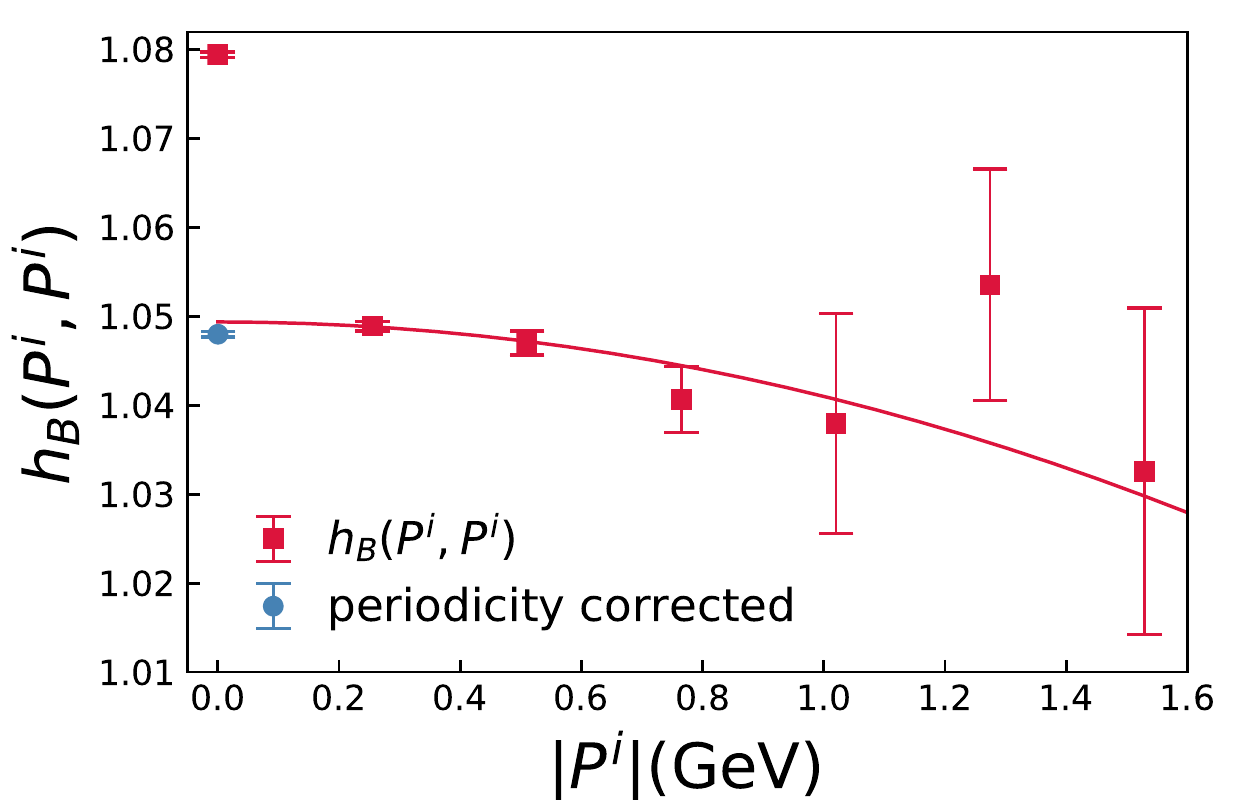}
\caption{The forward matrix elements $h_B(P^i,P^i)$. 
The $P_z^i$ dependence can be described by $h_B(P^i,P^i)=h_B^{ii}(P^i=0,P^i=0)+r(aP_z^i)^2$ shown as the line.\label{fig:zv}}
\end{figure}

\begin{figure}
\includegraphics[width=0.45\textwidth]{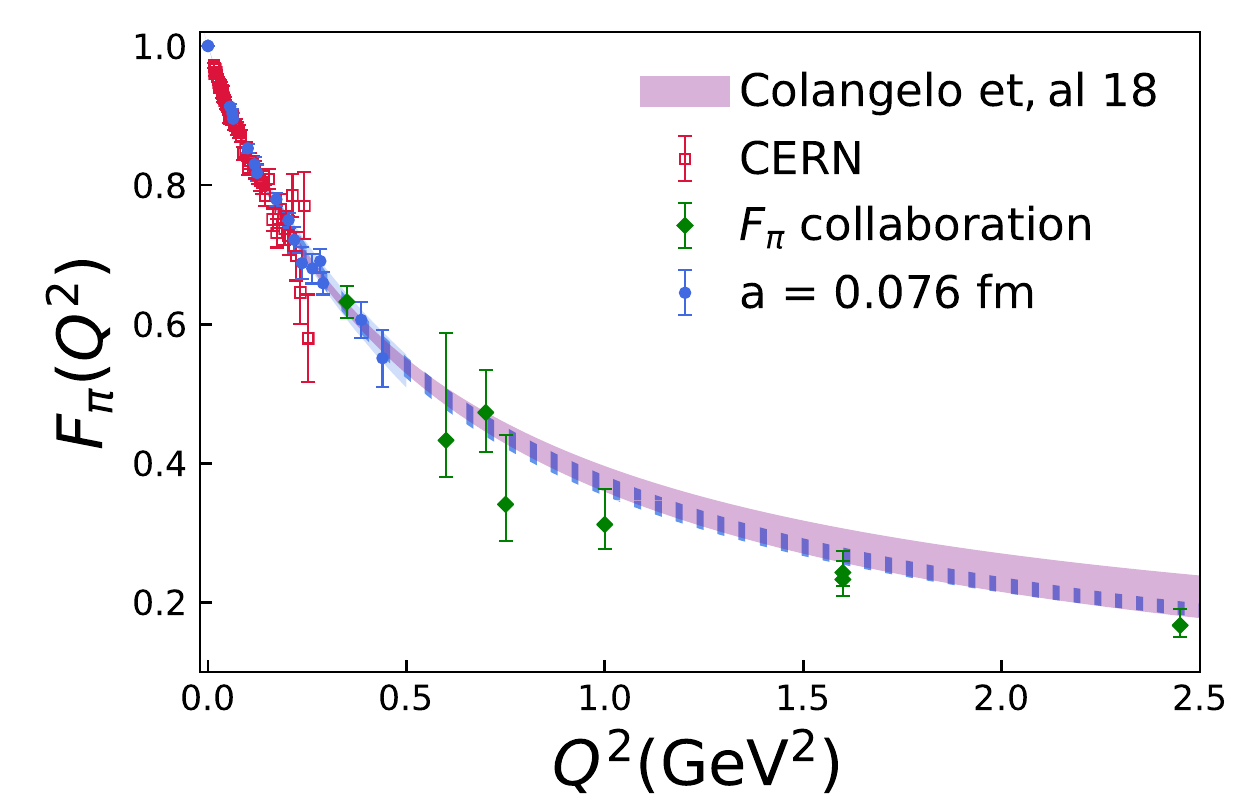}
\includegraphics[width=0.45\textwidth]{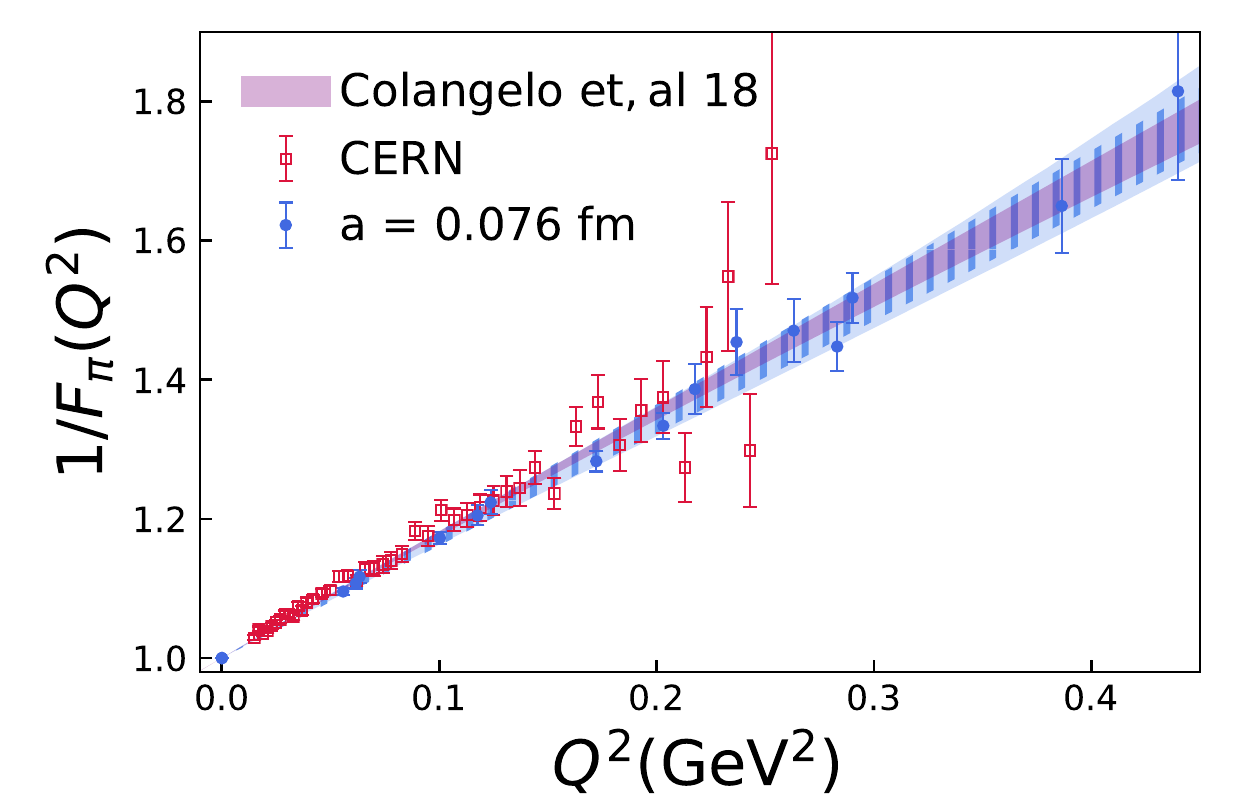}
\caption{Pion form factors (upper panel) and the inverse form factors (lower panel) derived from the a = 0.076 fm ($m_\pi=140$ MeV) ensemble (blue points), compared with the experiment data from CERN (red points) \cite{Amendolia:1986wj} and $F_{\pi}$ collaboration (green points) \cite{Huber:2008id}. The purle bands are the dispersive analysis results of experimental data from Ref.~\cite{Colangelo:2018mtw}, which also included form factors in time-like region. Our fit results of a = 0.076 fm data are shown as the blue bands, in which the filled band is from $z$-expansion fit and the dashed band is from monopole fit. The errors in this plot have included the systematic errors.\label{fig:FFrbm140}}
\end{figure}

To obtain the form factor from the bare form factor determined in the previous section it
needs to be multiplied by the vector current renormalization factor, $Z_V$. 
The simplest way to obtain this is to calculate the forward matrix element 
$h_B(P^i,P^i)=\langle 0 P^i | O | P^i 0\rangle = Z_V^{-1}$. 
However, one needs to keep in mind the wrap-around effect discussed in the previous section.
The other issue is cutoff dependence of $h_B(P^i,P^i)$ at large values of $P^i$.
In \fig{zv}, we show $h_B(P^i,P^i)$ for $a=0.076$ fm as a function of $P^i$.
In absence of discretization effects, $h_B(P^i,P^i)$ should be independent
of $P^i$ since after renormalization it gives the charge of the pion. 
In other words, $Z_V$ should not depend on the momentum of the external state.
Following
Ref.~\cite{Gao:2020ito}, we model the discretization effects using the form
$h_B(P^i,P^i)=h_B(P^i=0,P^i=0)+r(aP_z^i)^2$. As one can see from \fig{zv} this
form describes the data quite well, except for $P_i=0$. The anomalously large value of
$h_B(P^i,P^i)$ at $P_i=0$ is due to the wrap-around effects as discussed in the previous
section.  This means that $h_B(P^i,P^i)$ is contaminated by a small contribution
proportional to $e^{-aL_tm_\pi}$ mentioned in the previous section.
This contribution is also proportional to matrix elements containing two or more pion
states with the appropriate quantum numbers. Constraining such matrix elements is difficult
in practice. However, under some physically well-motivated assumptions it is possible
to estimate the corresponding contributions and remove them from $h_B(P^i,P^i)$ \cite{Gao:2020ito}.
Therefore, we follow the procedure 
explained in Appendix~A  of Ref.~\cite{Gao:2020ito} to remove this contribution from
the matrix element.
The corrected result for $h_B(P^i=0,P^i=0)$ is shown as the blue point in \fig{zv}
and is not very different from the result obtained by the fit.
Thus we understand the discretization effects in the forward matrix element $h_B(P^i,P^i)$.
We also calculated $Z_V$ for $a=0.076$ fm using RI-MOM scheme and obtained $Z_V=0.946(12)$ which agrees
with the results on $h_B(P^i=0,P^i=0)$ shown in \fig{zv} within errors.

From \fig{zv} we also see that the discretization errors are smaller than 1\% for $P^i_{z}<1$ GeV
, and are less than 2\% for $P^i_z<1.6$ GeV.
Since the discretization effects as functions of $P^i_z$ will be similar for off-forward matrix 
element it is convenient to obtain the renormalized pion form factor
by simply dividing $h_B(P^f,P^i)$ by $h_B(P^i,P^i)$. Then we have 
$F_\pi(Q^2=0)=1$ by construction and the discretization errors for large $P^i_z$ are removed.
We still may have discretization errors proportional to $(a Q)^2$. Assuming that these discretization
errors are similar to the $(a P_z^i)^2$ discretization errors we can neglect them. This is because
other sources of errors for the form factors are significantly larger for the considered $Q^2$ range as we will see below. We comment further on the cutoff dependence in the form-factor in \app{app1}.

In \fig{FFrbm140}, we show the renormalized pion form factors obtained for the $m_\pi$ = 140 MeV ensemble 
and compared to
the experimental data from CERN  \cite{Amendolia:1986wj}, as well as the results from $F_{\pi}$ collaboration \cite{Huber:2008id}. The purple bands are the dispersive analysis results of experimental data from Ref.~\cite{Colangelo:2018mtw}, which also included form factors in time-like region.
We see good agreement between the lattice results and the experimental data within the estimated error bars
at low $Q^2$. It is expected that at low $Q^2$, the pion form factors  can be described well
by a simple monopole Ansatz motivated by the Vector Meson Dominance (VMD) model \cite{OConnell:1995fwv}
\begin{equation}\label{eq:monopole}
F_\pi(Q^2)=\frac{1}{1+Q^2/M^2}.
\end{equation}
The monopole mass $M$ should be close to the $\rho$ meson mass.
Therefore, in \fig{FFrbm140} we show the inverse of the pion from factor, $1/F_\pi(Q^2)$, as a function
of $Q^2$. We see that in the studied range of $Q^2$ the inverse form factor can be roughly described by a linear function up to $Q^2=0.4$ GeV within the errors, as expected from monople form. The monopole fit of the lattice data (dashed band in \fig{FFrbm140}) extended to higher $Q^2$ also agrees with the pion form factor obtained by $F_{\pi}$ collaboration \cite{Huber:2008id}, possibly indicating that the monopole form
may work in an extended range of $Q^2$ within the current precision.

At very low $Q^2$, the pion form factor can be characterized in terms of the pion charge radius
\begin{equation}\label{eq:FFradius}
r_\pi^2=-6\frac{dF_\pi(Q^2)}{dQ^2}|_{Q^2=0}.
\end{equation}
As mentioned in the introduction, the pion charge radius is very sensitive to the quark mass, and it is clearly seen in the lattice calculations. In fact, it appears to be challenging
to obtain the correct pion charge radius from the lattice results
\cite{Brommel:2006ww,Frezzotti:2008dr,Aoki:2009qn,Brandt:2013dua,Alexandrou:2017blh,Bonnet:2004fr,Boyle:2008yd,Nguyen:2011ek,Fukaya:2014jka,Aoki:2015pba,Feng:2019geu,Wang:2020nbf,Koponen:2015tkr}.
The lattice calculations at the unphysical quark masses lead to smaller pion charge
radius than the experimental results. 
If the monopole form (\ref{eq:monopole}) could describe the pion form factor for all $Q^2$
the pion charge radius would be related to the monopole mass as 
\begin{equation}
r_\pi=\frac{\sqrt{6}}{M}.
\end{equation}

\begin{figure}
\includegraphics[width=0.45\textwidth]{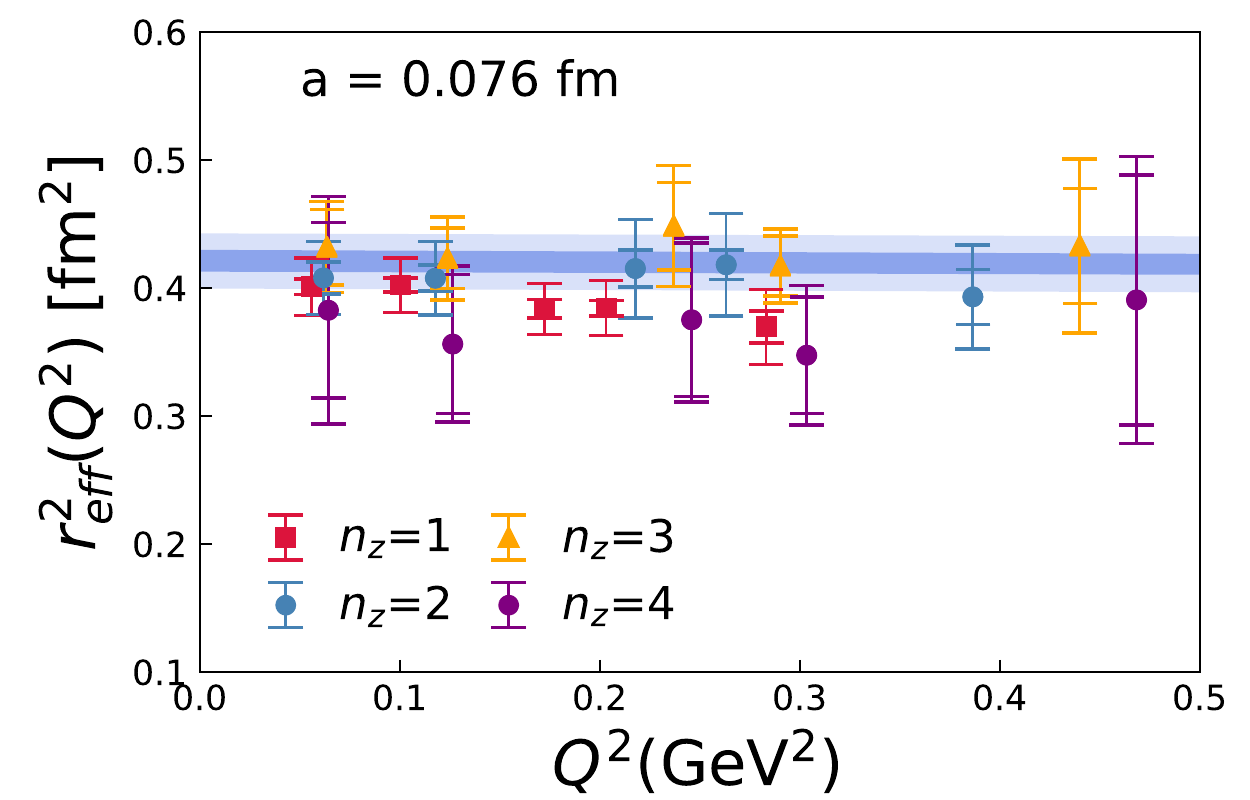}
\includegraphics[width=0.45\textwidth]{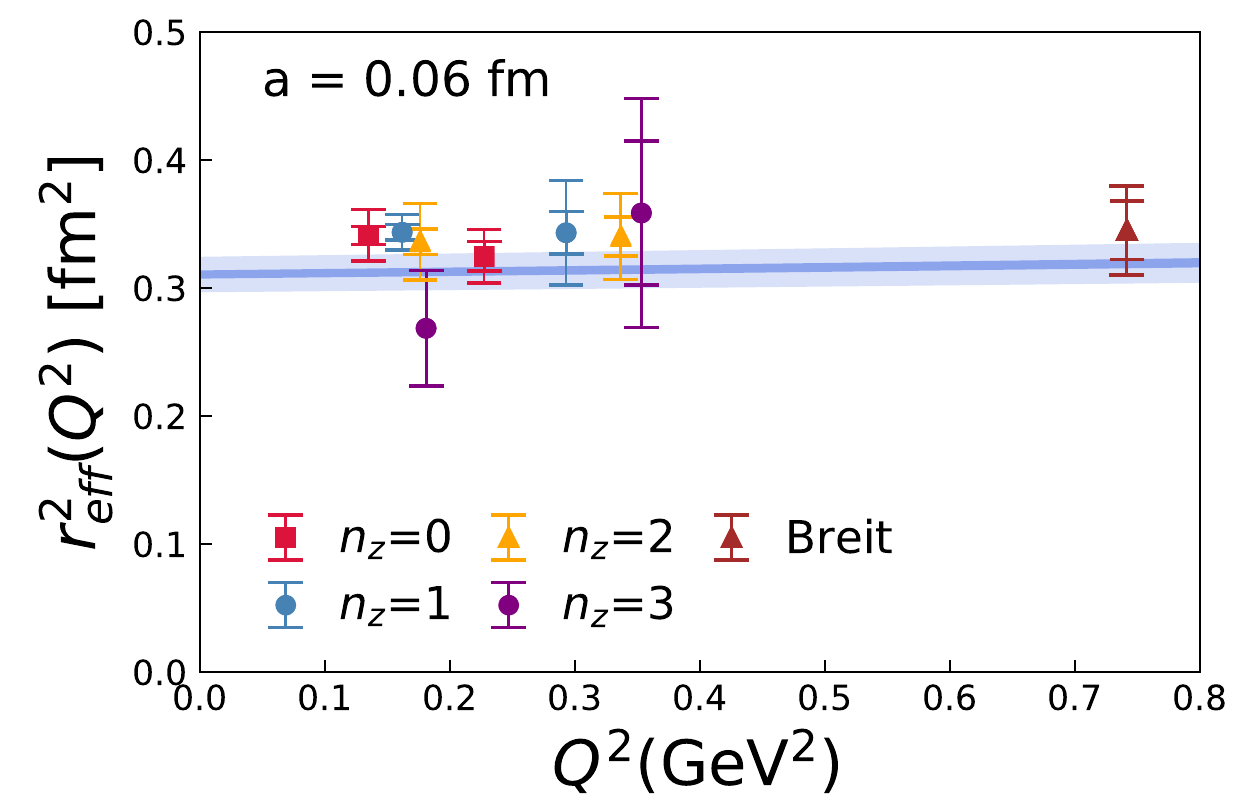}
\includegraphics[width=0.45\textwidth]{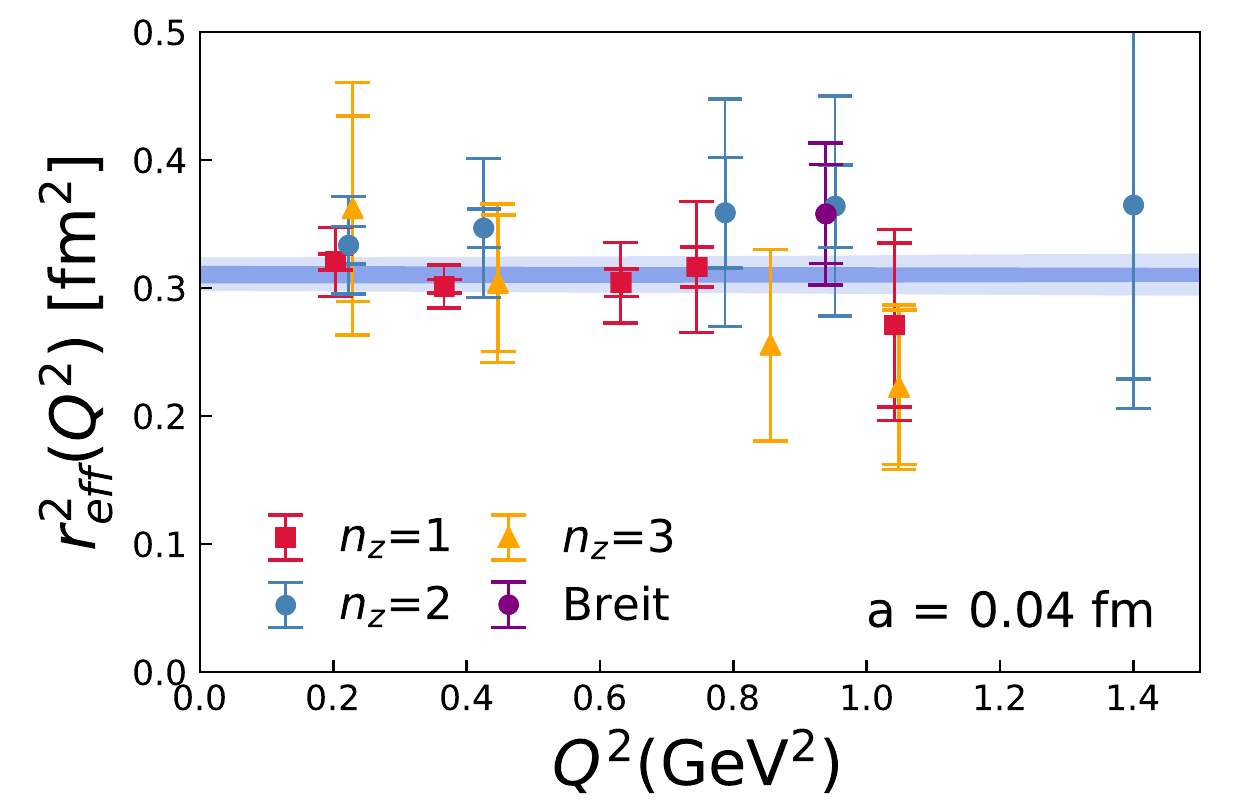}
\caption{The effective radius as a function of $Q^2$. The smaller error bars are the statistical
errors, while the larger error bars also include the systematic errors. 
We show results for $a=0.076$ fm (top panel), $a=0.06$ fm (middle panel) and $a=0.04$ fm (bottom panel).
The blue band is constructed by solving \Eq{FFradius} using $z$-expansion fit results as a function of $Q^2$.\label{fig:Effradius}}
\end{figure}

It is convenient to represent the form factors in terms of the effective charge radius defined as \cite{Brommel:2006ww}
\begin{equation}\label{eq:Effradius}
r^2_{eff}(Q^2)=\frac{6(1/F_\pi (Q^2)-1)}{Q^2}.
\end{equation}
In \fig{Effradius} we show the effective radius for $a=0.076$ fm ensemble
as well as for the two finer ensembles with $m_{\pi}^{val}=300$ MeV.
We see from the figure that $r^2_{eff}$ is roughly constant as a function of $Q^2$ for all three lattice spacings. For the smallest lattice spacing, $a=0.04$ fm the results on the effective radius are $Q^2$-independent for $Q^2$ as high as $1.4$ $\rm GeV^2$. This is consistent with earlier findings \cite{Brommel:2006ww}. We also clearly see the quark mass dependence of $r^2_{eff}$. The effective radius is
smaller for the heavier pion mass as expected. Comparing the results at $a=0.06$ fm and $a=0.04$ fm we see no clear lattice spacing dependence of $r^2_{eff}$. Therefore, we conclude that for $a=0.06$ fm the discretization errors for the pion form factor are smaller than the estimated lattice errors in the range of $Q^2$ studied by us. Finally, for the two finer lattices we also show the results from the calculations using Breit frame, which agree with the non-Breit frame results.

\begin{figure}
\centering
        \includegraphics[width=0.45\textwidth]{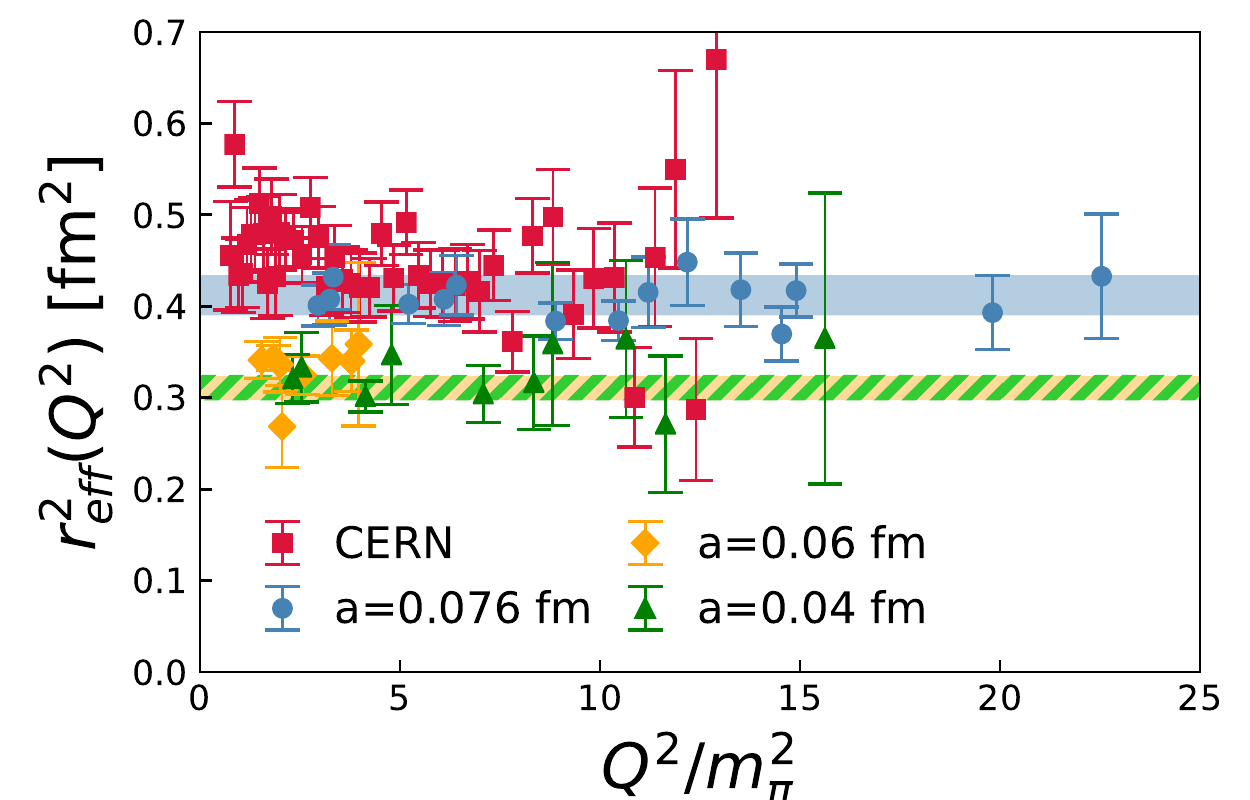}
        \caption{The comparison of effective radius between CERN and our lattice data as a function of $Q^2/m_\pi^2$. The bands are the z expansion fit results of lattice data (blue, green and orange). \label{fig:ComExp}}
\end{figure}

While the monopole Ansatz seems to describe the pion form factor well
and was used to obtain the pion charge radius in the past (see e.g. Ref. \cite{Brommel:2006ww})
there is no strong theoretical reason why it should describe the pion form factor. Therefore,
one has 
to consider an alternative and more flexible parameterization of the pion form factor.
An alternative way to fit the form factors is the model independent method called the $z$-expansion \cite{Lee:2015jqa}.
Here the form factor is written as
\begin{equation}\label{eq:zfit}
\begin{aligned}
&F_\pi(Q^2)=\sum_{k=0}^{k_{max}}a_kz^k\\
&z(t,t_{\textup{cut}},t_0)=\frac{\sqrt{t_{\textup{cut}}-t}-\sqrt{t_{\textup{cut}}-t_0}}{\sqrt{t_{\textup{cut}}-t}+\sqrt{t_{\textup{cut}}-t_0}}
\end{aligned}
\end{equation}
where $t=-Q^2$, $a_k$ are the fit parameters with con-
strain condition $F_\pi(Q^2=0)=1$, and $t_{\textup{cut}}=4m_\pi^2$ is the two-pion production threshold. Furthermore, $t_0$ is chosen to be the optimal value $t^{\rm opt}_0(Q^2_{\textup{max}})=t_{\textup{cut}}(1-\sqrt{1+Q^2_{\textup{max}}/t_{\textup{cut}}})$ to minimize the maximum value of $|z|$, with $Q^2_{\textup{max}}$ the maximum $Q^2$ used for the fit. In the timelike region near the two pion threshold, the leading singularity of form factor should be proportional to $(4 m_{\pi}^2-t)^{3/2}$ due to the P-wave nature of the $\pi - \pi$ scattering \cite{Colangelo:2018mtw,Leutwyler:2002hm,Colangelo:2003yw}, which leads to the additional constraint $\sum_{k=1}^{k_{max}}(-1)^kka_k=0$.
We use AIC model selection rules to determine $k_{max}$, 
which are 2 for a = 0.06 fm, and 3 for a = 0.04, 0.076 fm data and for the $Q^2$ under consideration.
The $z$ expansion results are also shown in \fig{FFrbm140} and appear to overlap with the monopole fit, but for larger $Q^2$ it has larger errors. We also show the fits with the $z$-expansion in \fig{Effradius}. From this figure we see that this fit works well also for the valence pion mass of $300$ MeV and naturally reproduces little $Q^2$ dependence of the effective radii. To better understand the quark mass dependence of the pion form factor as well to facilitate the comparison with the experimental results, in \fig{ComExp}  we show all the results for the pion form factor
in terms of the effective radius $r_{eff}(Q^2)$. We see that the effective radius obtained for the physical
pion mass is clearly larger than the one obtained for $m_{\pi}^{val}=300$ MeV and is much closer to the
CERN data.
Furthermore, the fits of $r_{eff}$
for  $m_{\pi}^{val}=300$ MeV for the two lattice spacings agree within errors.
While the individual lattice data and the CERN data appear to agree within errors we also see from
the figure that there is a tendency for the CERN data to lie higher than the lattice data. This leads to a slight difference in the pion charge
radius as discussed below.

\begin{figure}
\centering
        \includegraphics[width=0.4\textwidth]{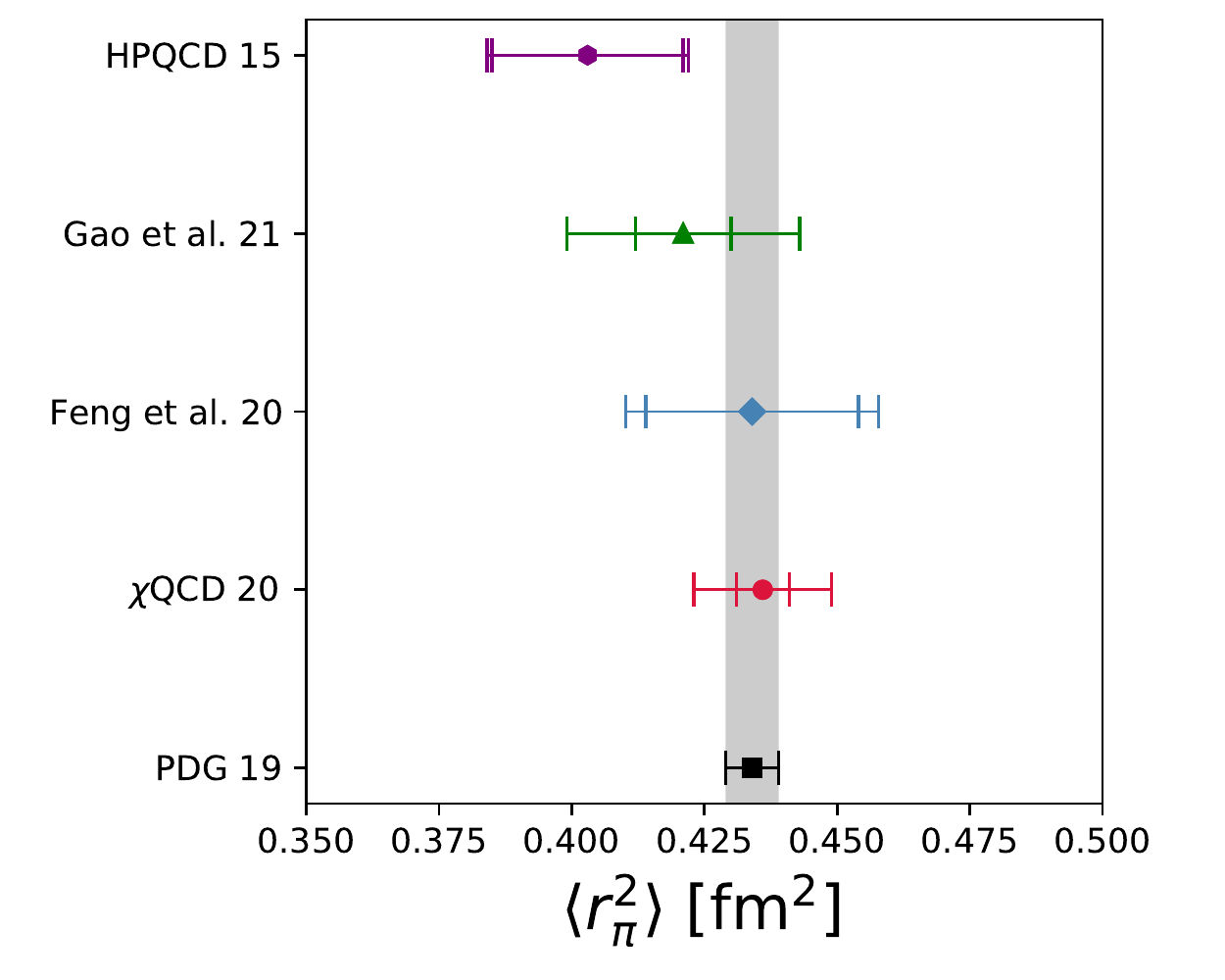}
        \caption{The comparison of pion radius between determination from lattice QCD at physical point and the PDG value. The shown lattice results come from this work (green), HPQCD\cite{Koponen:2015tkr} (purple), Feng et al\cite{Feng:2019geu} (blue), $\chi$QCD\cite{Wang:2020nbf} (red). \label{fig:ComRadius}}
\end{figure}

The pion charge radius can be derived from $z$-expansion fit results using \Eq{FFradius}, which are summarized in Table \ref{tb:radius} for the three lattice spacings used in this work. We also discuss the radius obtained from the monopole fit for comparison in \app{app3}.  As expected the calculations for the heavier quark mass give smaller pion charge radius. Since the $z$-expansion provides a model independent way to obtain the pion charge radius, for our final estimate of the pion charge radius at the physical point we take
the result from the $z$-expansion fit: 
\begin{equation}
\langle r_{\pi}^2 \rangle=0.42(2)~{\rm fm^2},
\end{equation}
where we added the statistical and systematic errors (defined by the difference between the results from 2-state and 3-state fit of matrix elements) in quadrature. This result is consistent the pion charge radius quoted by Particle Data Group (PDG), $\langle r_{\pi}^2\rangle_{\rm PDG}=0.434(5)~{\rm fm}^2$ \cite{Zyla:2020zbs}, which is averaged from determination from t-channel $\pi e {\rightarrow} \pi e$ scattering data \cite{Dally:1982zk,Amendolia:1986wj,GoughEschrich:2001ji} and s-channel $e^{+} e^{-} {\rightarrow} \pi^{+} \pi^{-}$ data sets \cite{Ananthanarayan:2017efc,Colangelo:2018mtw}. The HPQCD determination that uses HISQ action both in the sea and the valence sectors of
$(2+1+1)$-flavor
QCD is $\langle r_{\pi}^2\rangle=0.403(18)(6)~{\rm fm}^2$ \cite{Koponen:2015tkr}. The most precise lattice determination
of the pion charge radius in 2+1 flavor QCD using overlap
action in the valence sector and domain wall action in the sea sector has 
$\langle r_{\pi}^2 \rangle=0.436(5)(12)~{\rm fm}^2$ \cite{Wang:2020nbf}.
The 2+1 flavor domain wall calculation gives $\langle r_{\pi}^2\rangle=0.434(20)(13)~{\rm fm}^2$ \cite{Feng:2019geu}. 
Finally, the other 2+1 flavor
lattice determinations of the pion charge radius have significantly larger errors \cite{Fukaya:2014jka,Aoki:2015pba}. We summarize the comparison in \fig{ComRadius}.

\begin{table}
\centering
\begin{tabular}{|c|c|c|c|c|c|c|}
\hline
\hline 
Data&$n_z$&$\langle r_{\pi}^2 \rangle$ [$\rm{fm^2}$]\\
\hline 
a=0.076fm&[1,3]&0.421(9)(20)\\
\hline 
a=0.06fm&[0,3]&0.311(3)(13)\\
\hline 
a=0.04fm&[1,3]&0.311(8)(11)\\
\hline
\hline
\end{tabular}
\caption{The charge radius computed from $z$-expansion fit. The first error is statistical, while
the second error is systematic.
}\label{tb:radius}
\end{table}

\section{Conclusions}
In this paper we studied the pion form factor in 2+1 flavor lattice QCD
using three lattices spacings $a=0.076$, $a=0.06$ and $a=0.04$ fm.
The calculations on the coarsest lattice have been performed with
the physical value of the quark masses, while for the finer two lattices
the valence pion mass was $300$ MeV. We have found that the pion form factor
is very sensitive to the quark mass, as expected. We showed that lattice discretization
effects are quite small for lattice spacings smaller than $0.06$ fm. 
For the physical quark masses our lattice results on the pion form factor appear
to agree with the experimental determinations.
Unlike other lattice studies we also considered highly boosted pions in the initial
state using momentum boosted Gaussian sources. 
In addition we performed calculations 
also in the Breit frame. We demonstrated that the calculations of the pion
form factor performed at different momenta 
of the pion as well as in the Breit frame give consistent results.
This is very important for extending the calculations to pion GPDs.

An important outcome of our analysis is that the monopole Ansatz can describe
the pion form factor in large range of $Q^2$, up to $Q^2=1.4~{\rm GeV}^2$. 
In the future
it will be important to extend the calculations to even higher momentum transfer
given the experimental efforts in Jlab and EIC.
To do this we should use boosted sources that also depend on the value
of $Q^2$. At present the momentum boost was optimized only according 
to the pion momentum in the initial state.

From the low $Q^2$ dependence of the pion form factor we determined the pion
charge radius, which is one sigma lower that the experimental result. We 
speculated, whether this is due to the effect of partial quenching. To
fully resolve this issue calculations at smaller lattice spacing with
the physical value of the pion masses are needed.

\section*{Aknowledgements}
We thank Gilberto Colangelo, Martin Hoferichter, Peter Stoffer for
their comments to the earlier version of the manuscript.
This material is based upon work supported by: 
(i) The U.S. Department of Energy, Office of Science, 
Office of Nuclear Physics through the Contract Nos. DE- SC0012704 and DE-AC02-06CH11357; 
(ii) The U.S. Department of Energy, 
Office of Science, Office of Nuclear Physics and Office
of Advanced Scientific Computing Research within the framework of Scientific Discovery through Advance 
Computing (ScIDAC) award Computing the Properties of Matter with Leadership Computing Resources;
(iii) X.G. is partially supported by the NSFC Grant Number 11890712. 
(iv) N.K. is supported by Jefferson Science Associates, 
LLC under U.S. DOE Contract No. DE- AC05-06OR23177 and in part by U.S. DOE grant No. DE-FG02-04ER41302. 
(v) S.S. is supported by the National Science Foundation under CAREER Award PHY- 1847893 and by the RHIC Physics Fellow Program of
the RIKEN BNL Research Center.
(vi) This research used awards of computer time provided by the INCITE and ALCC programs at Oak Ridge Leadership Computing Facility, 
a DOE Office of Science User Facility operated under Contract No. DE-AC05-00OR22725. 
(vii) Computations for this work were carried out in part on facilities of the USQCD Collaboration, 
which are funded by the Office of Science of the U.S. Department of Energy.

\appendix 
\section{Discretization errors}\label{app:app1}
\begin{figure}
\includegraphics[width=0.45\textwidth]{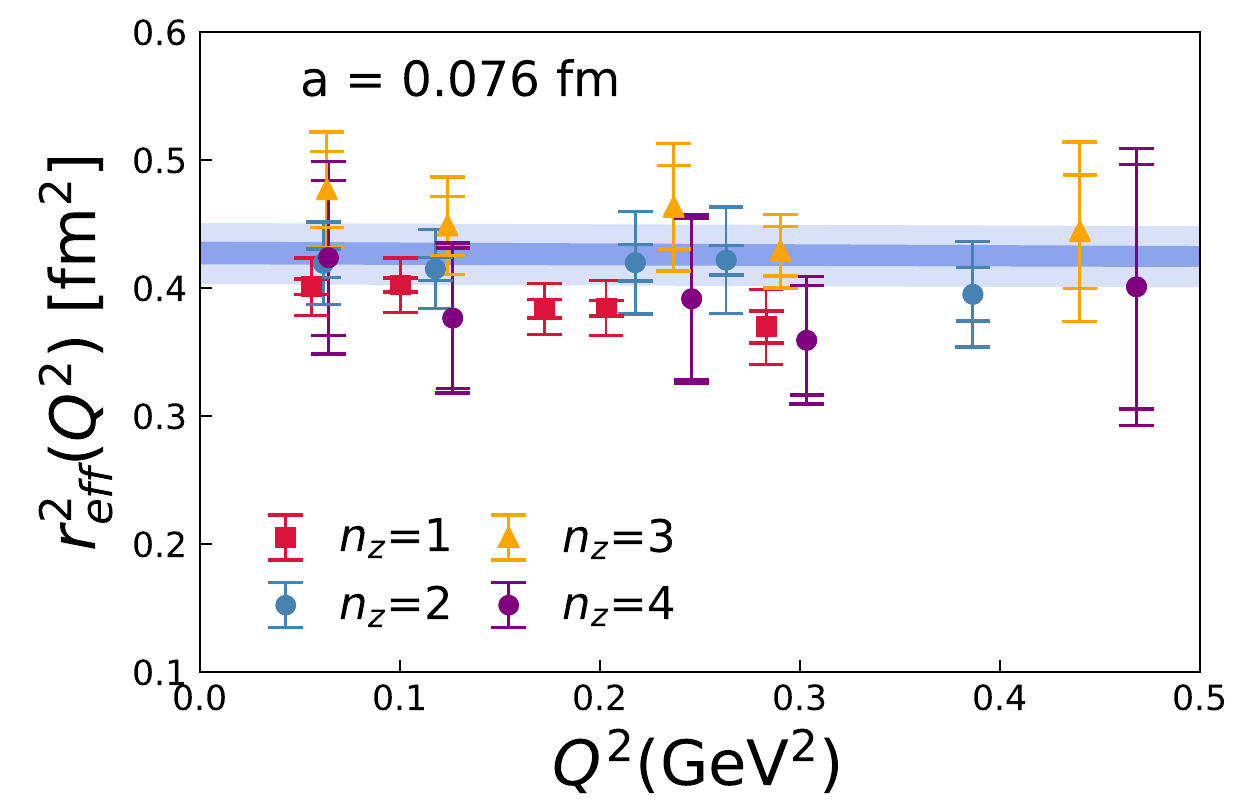}
\caption{Similar plot as \fig{Effradius} for $a=0.076$ fm ensemble but using constant $Z_V^{-1}$ for renormalization is shown.\label{fig:Effradiuspz1}}
\end{figure}

As is shown in \fig{zv}, there are $\lesssim 2\%$ discretization effects of $Z_V^{-1}(P^i)=h_B(P^i,P^i)$. We chose to divide $h_B(P^f,P^i)$ by $h_B(P^i,P^i)$ so that the renormalized pion form factors could reduce such effects. To estimate the impact of the discretization errors to the form factors as well as pion charge radius, instead we can renormalize the bare form factors $h_B(P^f,P^i)$ by a constant $Z_V^{-1}$ such as $Z_V^{-1}$(0.25~GeV) of a = 0.076 fm ensemble. The effective radius for a = 0.076 fm ensemble is shown in \fig{Effradiuspz1}, and in this case we estimate the charge radius from monopole fit and $z$-expansion fit as 0.406(6)(25) $\rm{fm^2}$ and 0.427(10)(22) $\rm{fm^2}$, which shift $\lesssim 2\%$ but are consistent with the estimates in \tb{radius}.
\section{Form factors from plateau fit}\label{app:app2}
\begin{figure*}
\includegraphics[width=0.3\textwidth]{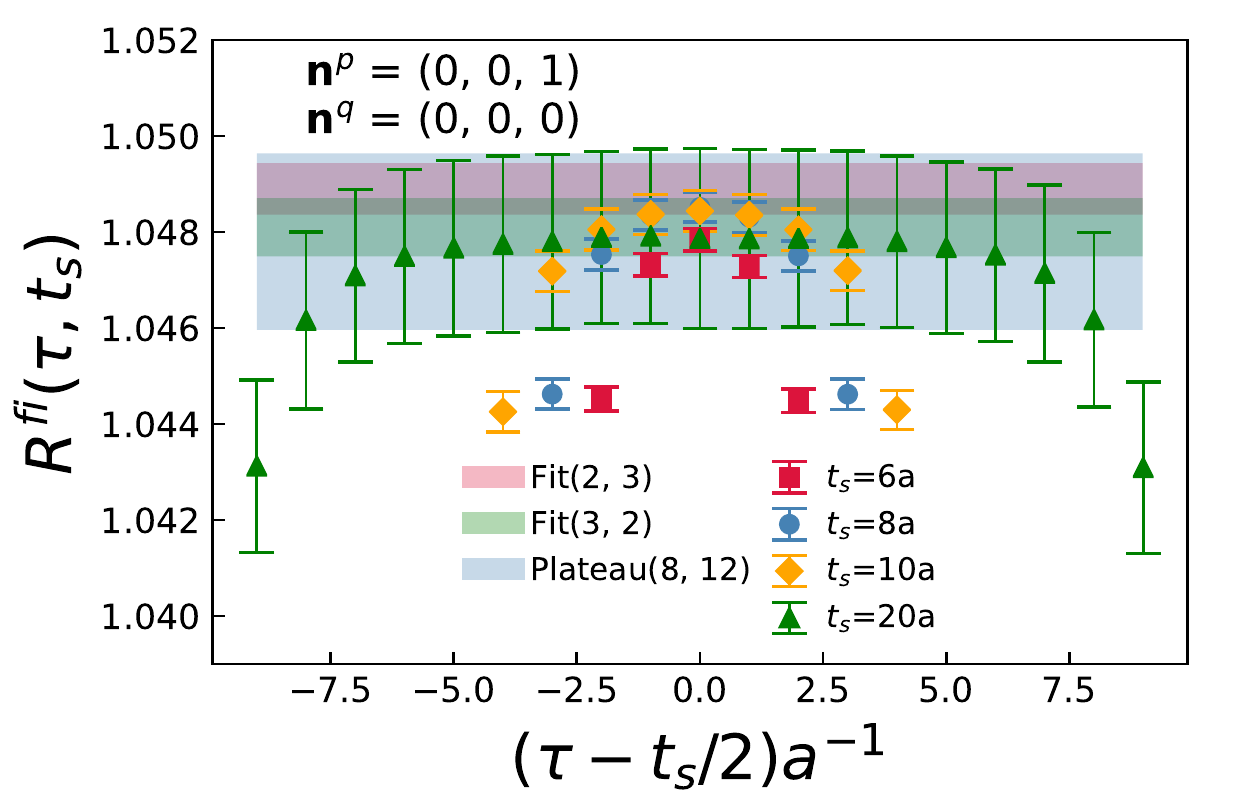}
\includegraphics[width=0.3\textwidth]{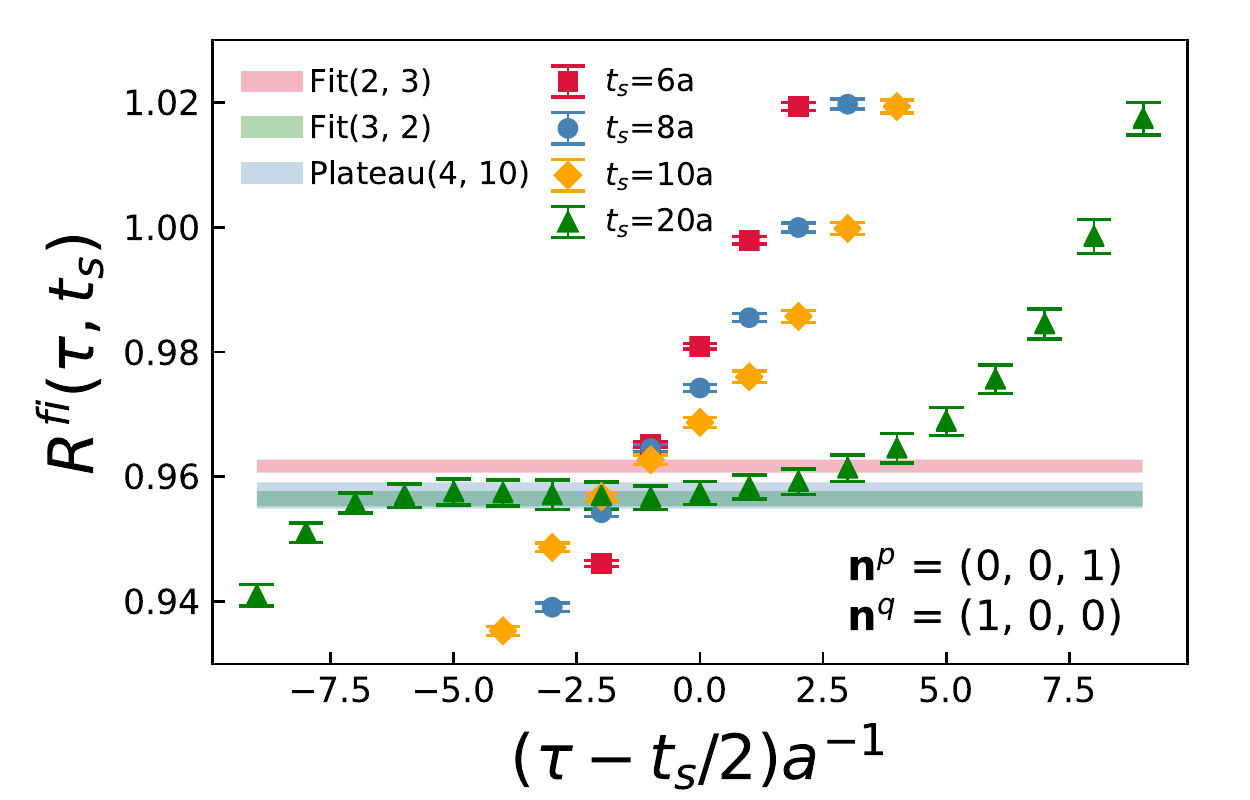}
\includegraphics[width=0.3\textwidth]{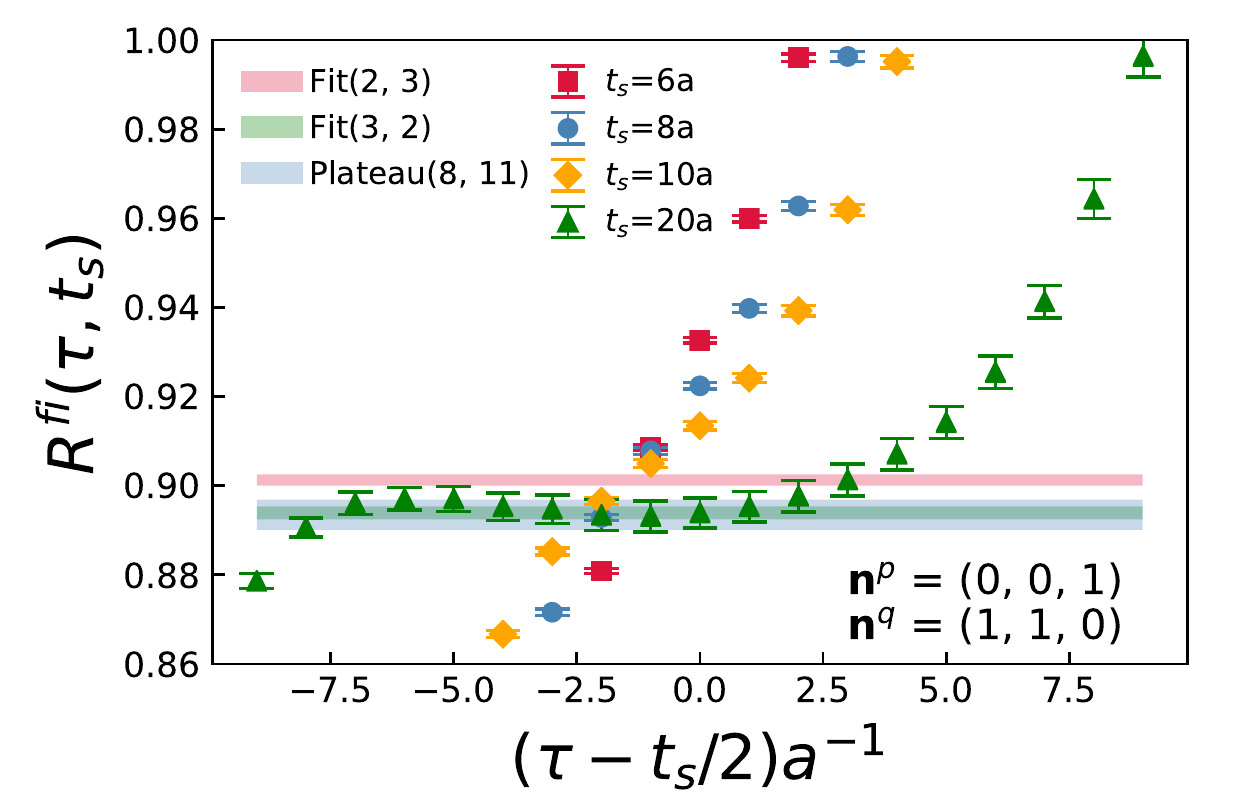}
\includegraphics[width=0.3\textwidth]{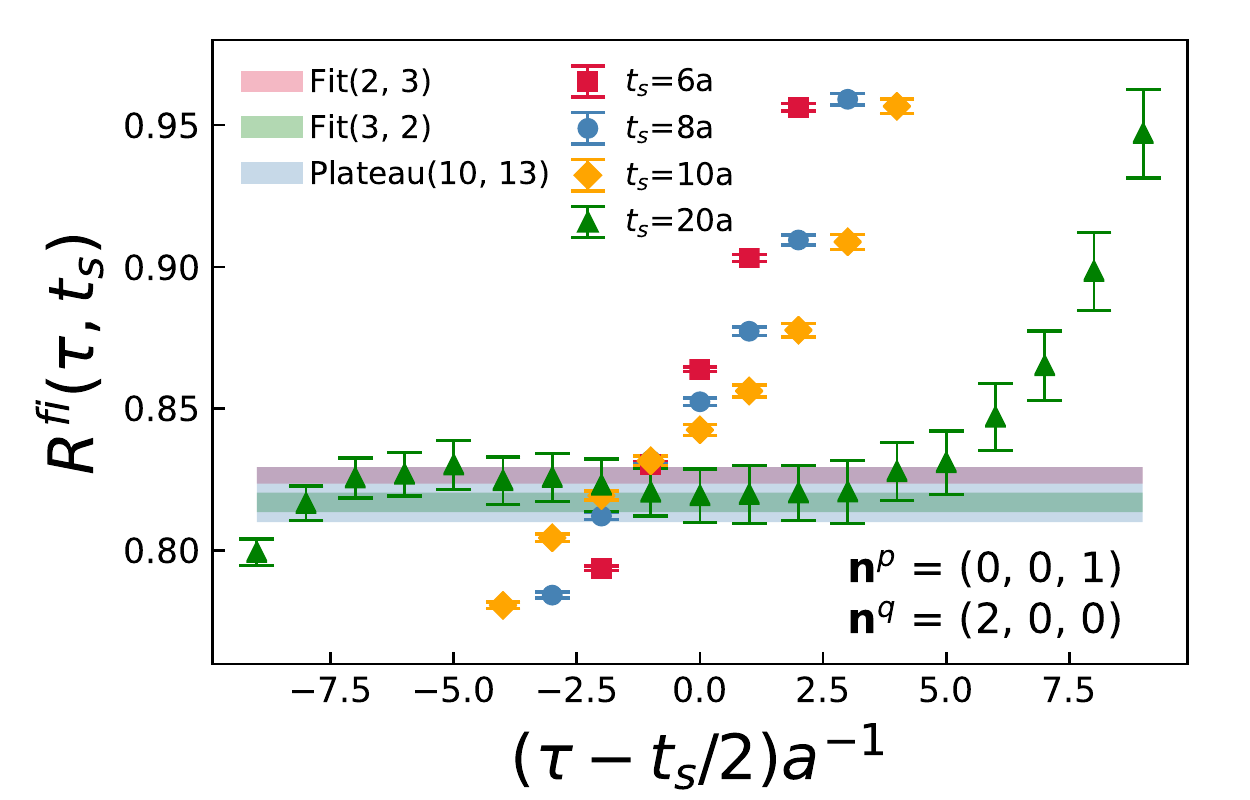}
\includegraphics[width=0.3\textwidth]{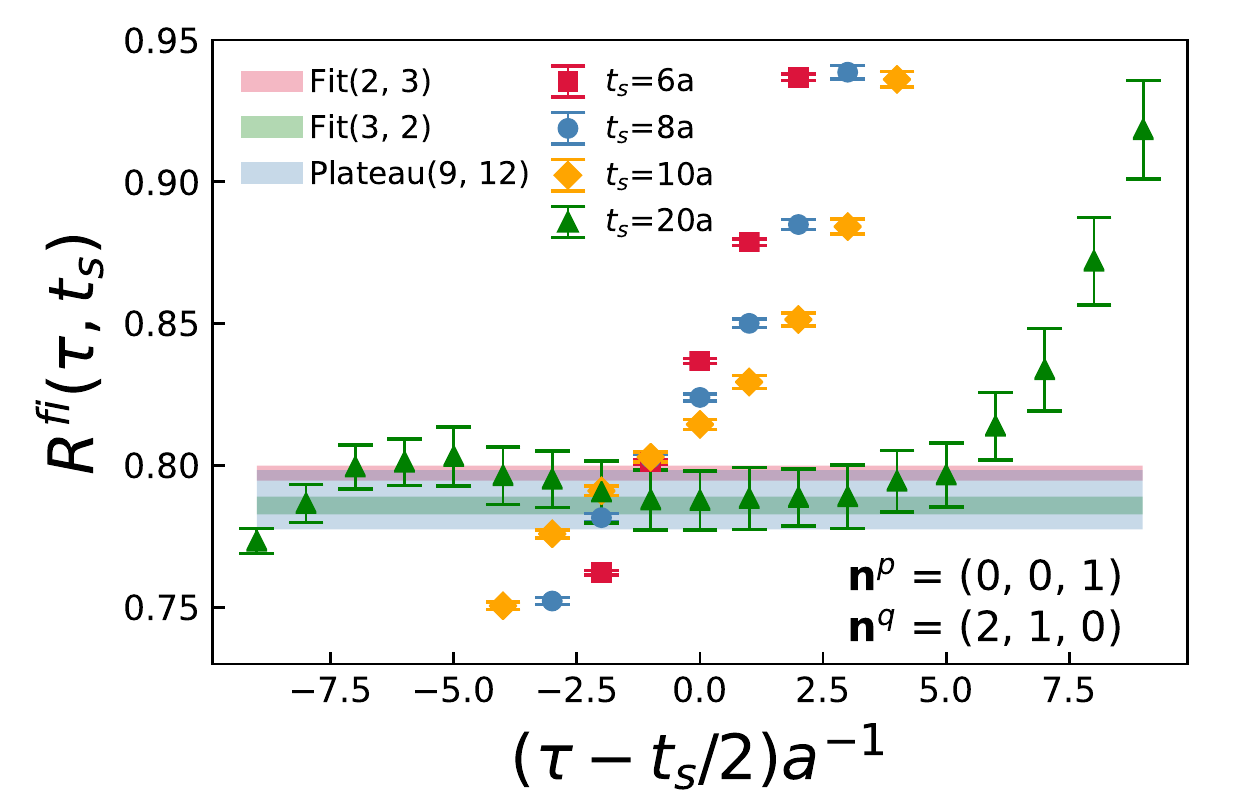}
\includegraphics[width=0.3\textwidth]{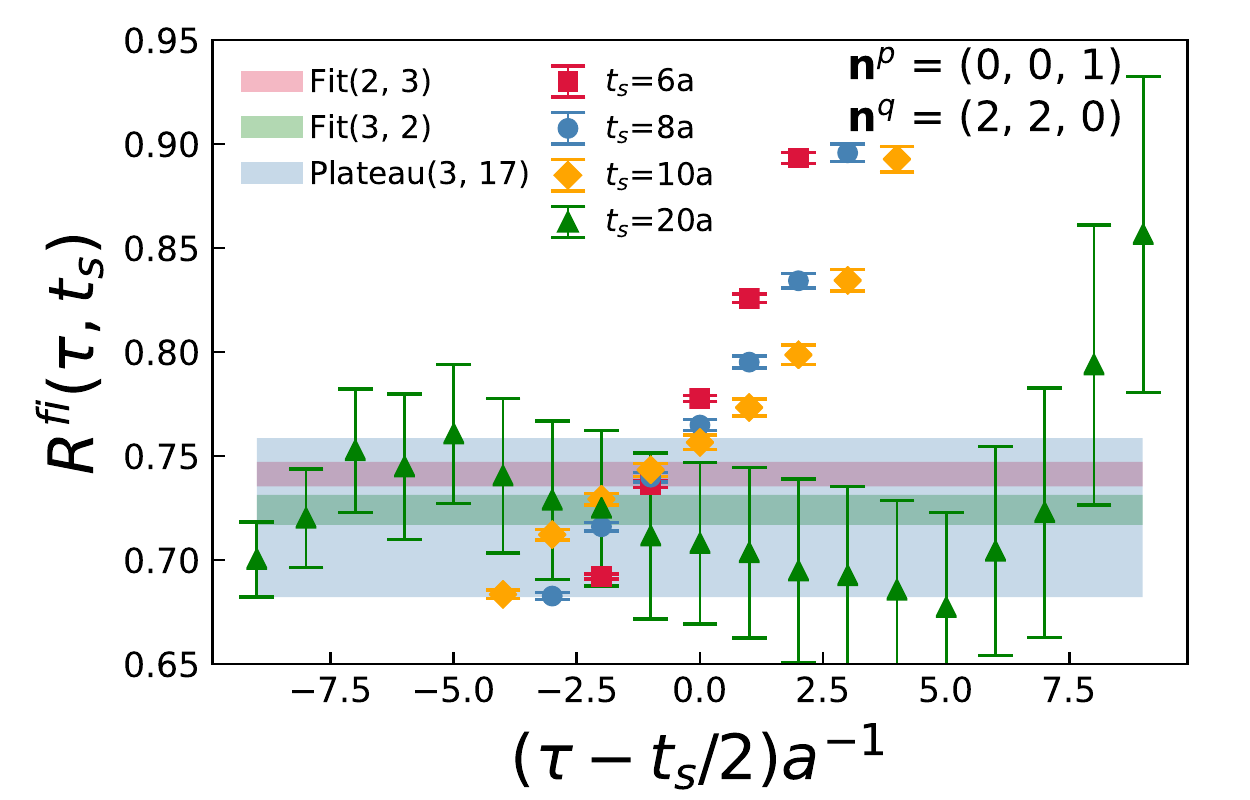}
\caption{$R^{fi}(\tau,t_s)$ for $\mathbf{n^p}_i = (0,0,1)$ with $\mathbf{n^q}=(0,0,0),~(1,0,0),~(1,1,0),~(2,0,0),~(2,1,0),~(2,2,0)$ of physical ensemble are shown. The bands are the estimated bare matrix elements from $\textup{Fit}(N_{state},n_{sk})$ using $t_s$ = 6a, 8a, 10a and Plateau($\tau_{\rm min},~\tau_{\rm max}$) using $t_s$ = 20a. The errors are estimated using bootstrap method. \label{fig:plateaufit}}
\end{figure*}

\begin{figure*}
\includegraphics[width=0.3\textwidth]{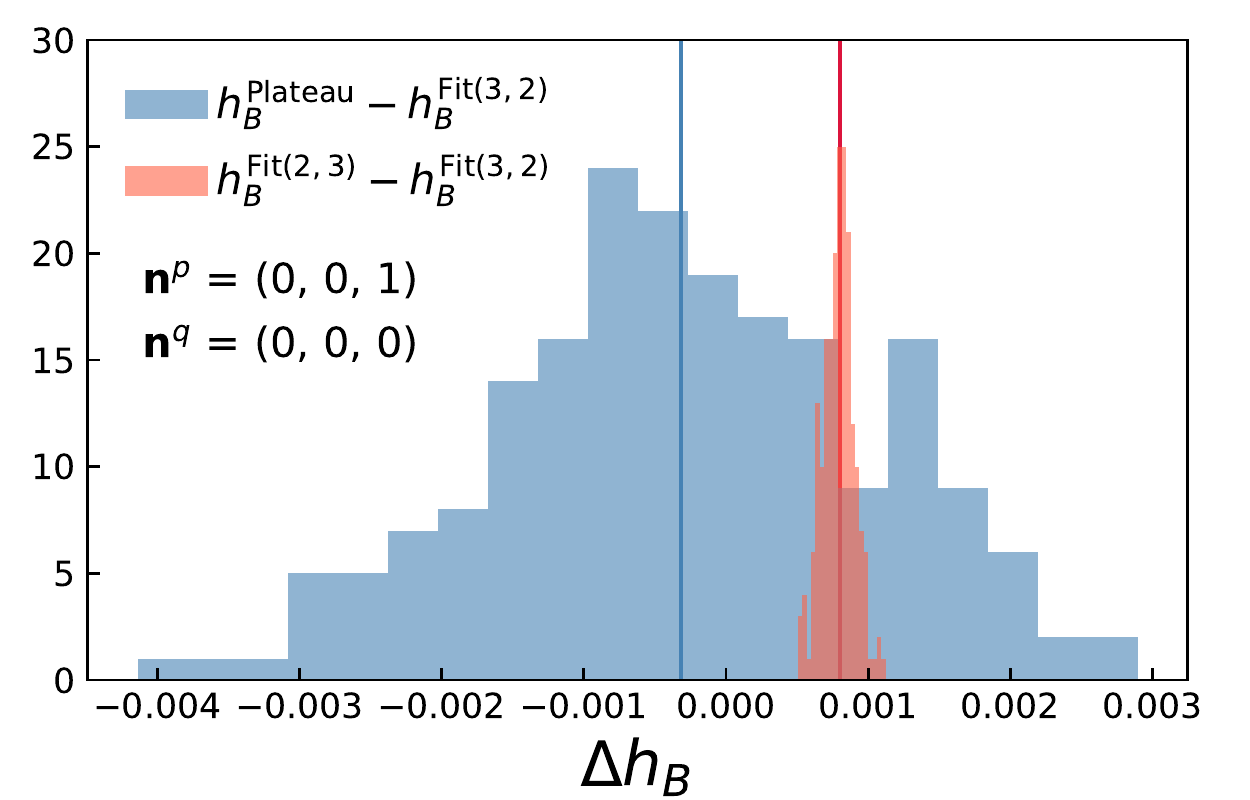}
\includegraphics[width=0.3\textwidth]{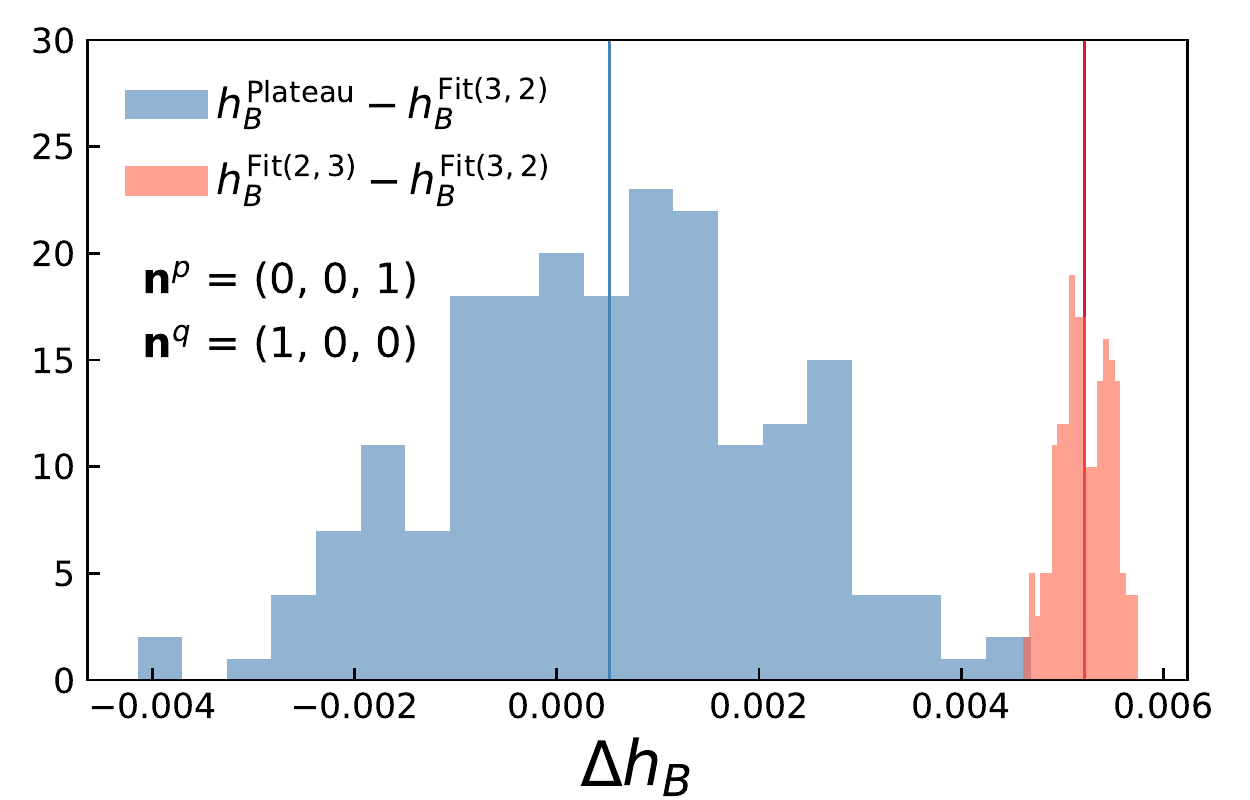}
\includegraphics[width=0.3\textwidth]{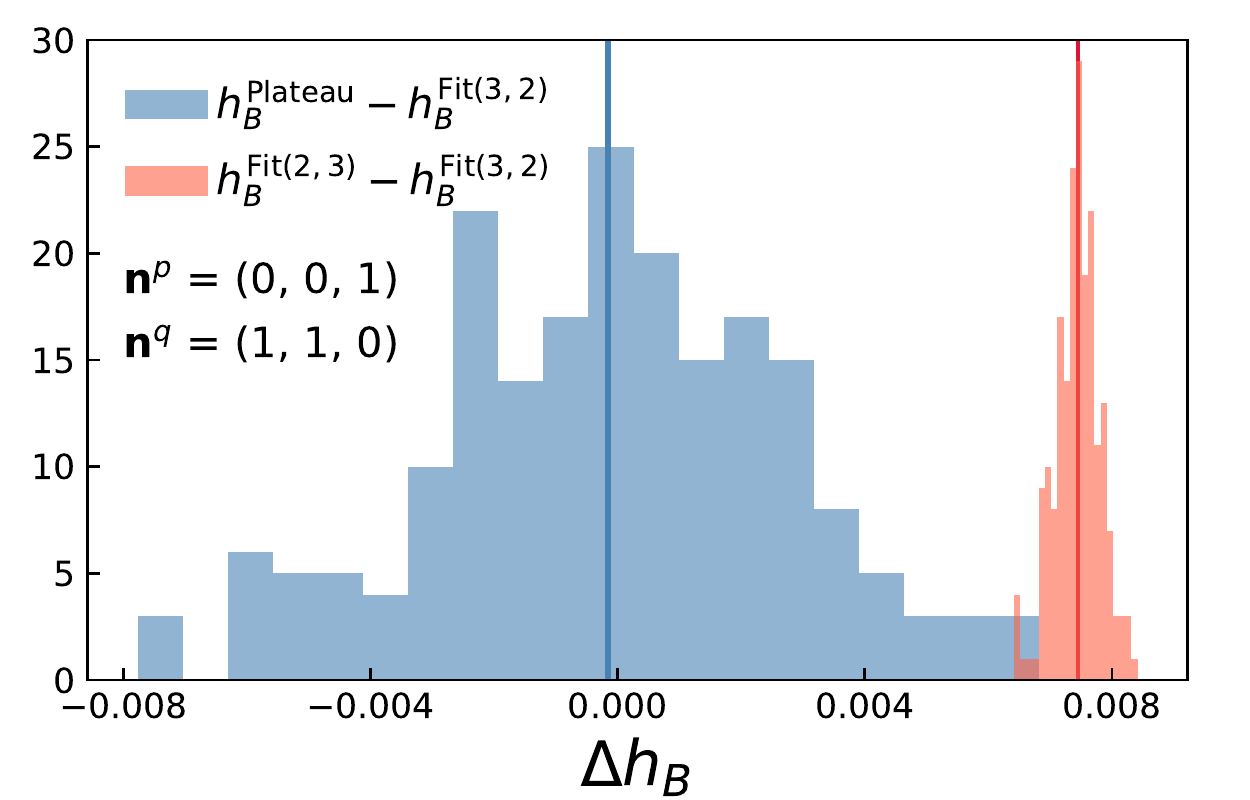}
\includegraphics[width=0.3\textwidth]{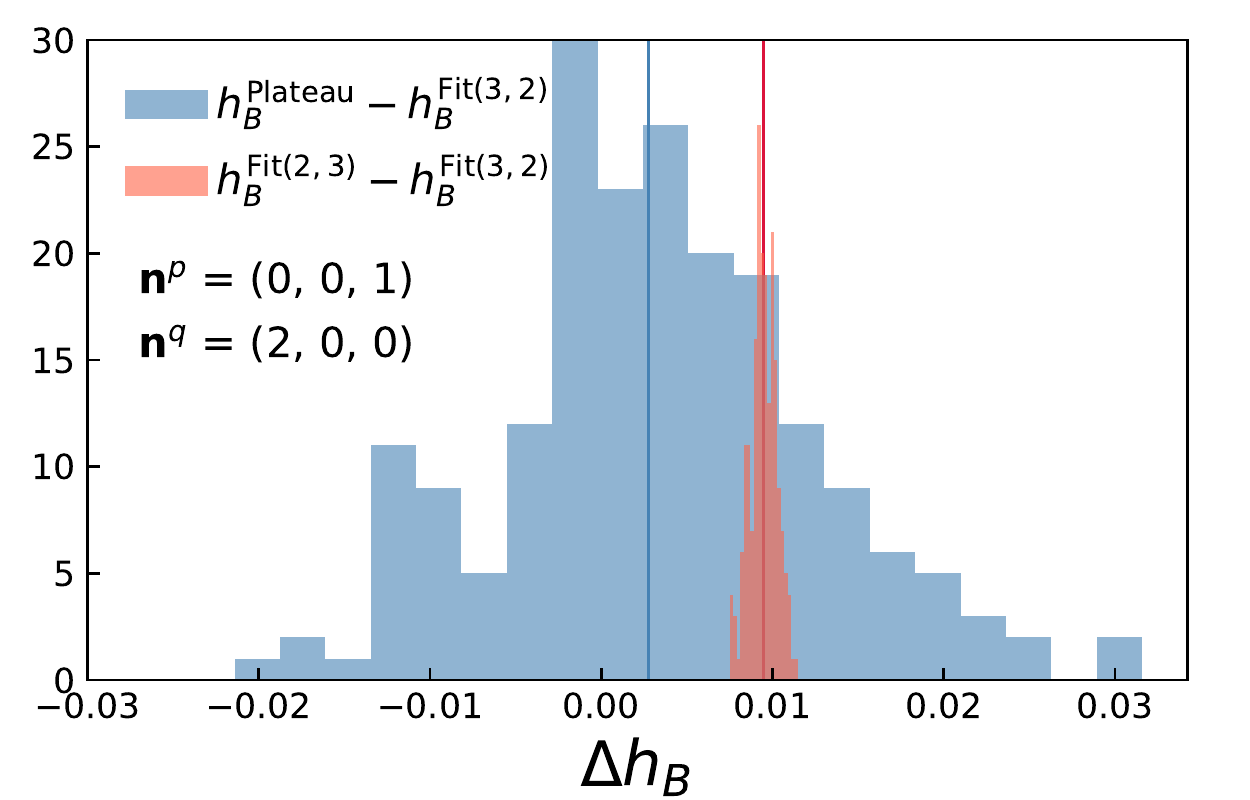}
\includegraphics[width=0.3\textwidth]{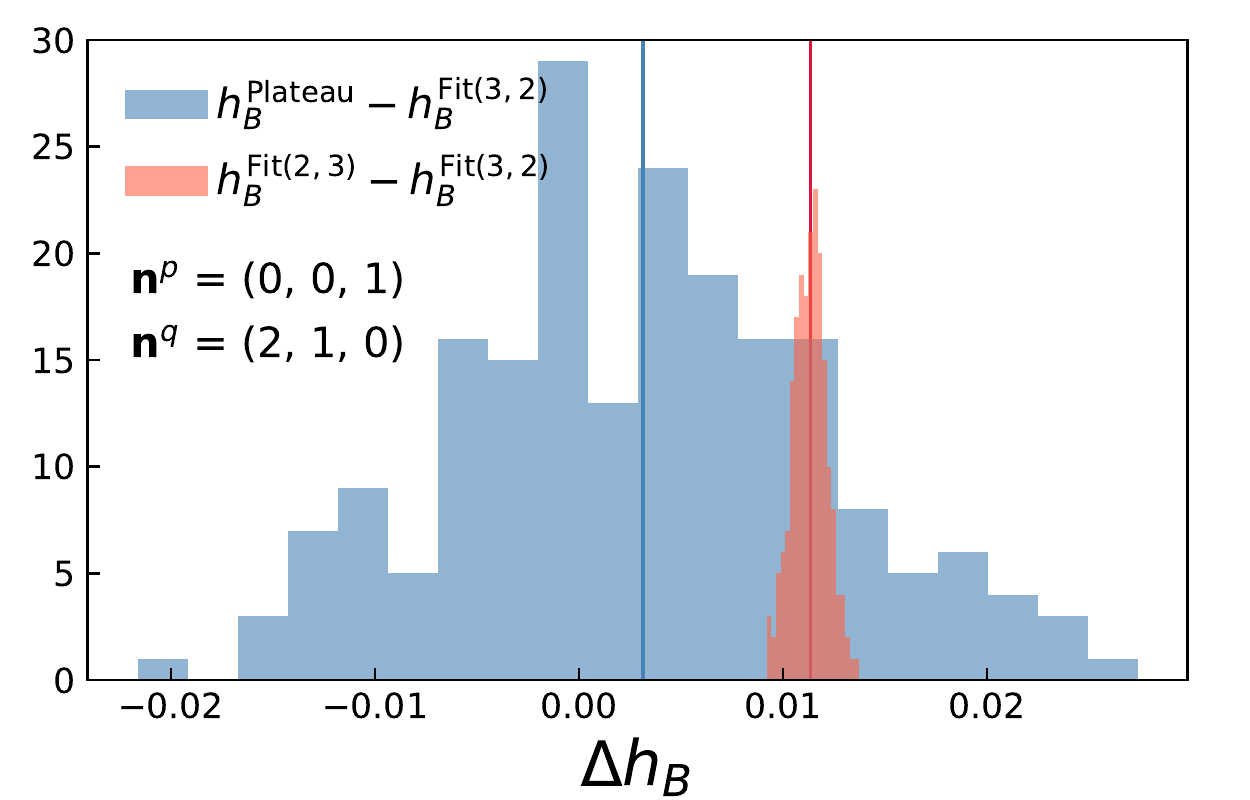}
\includegraphics[width=0.3\textwidth]{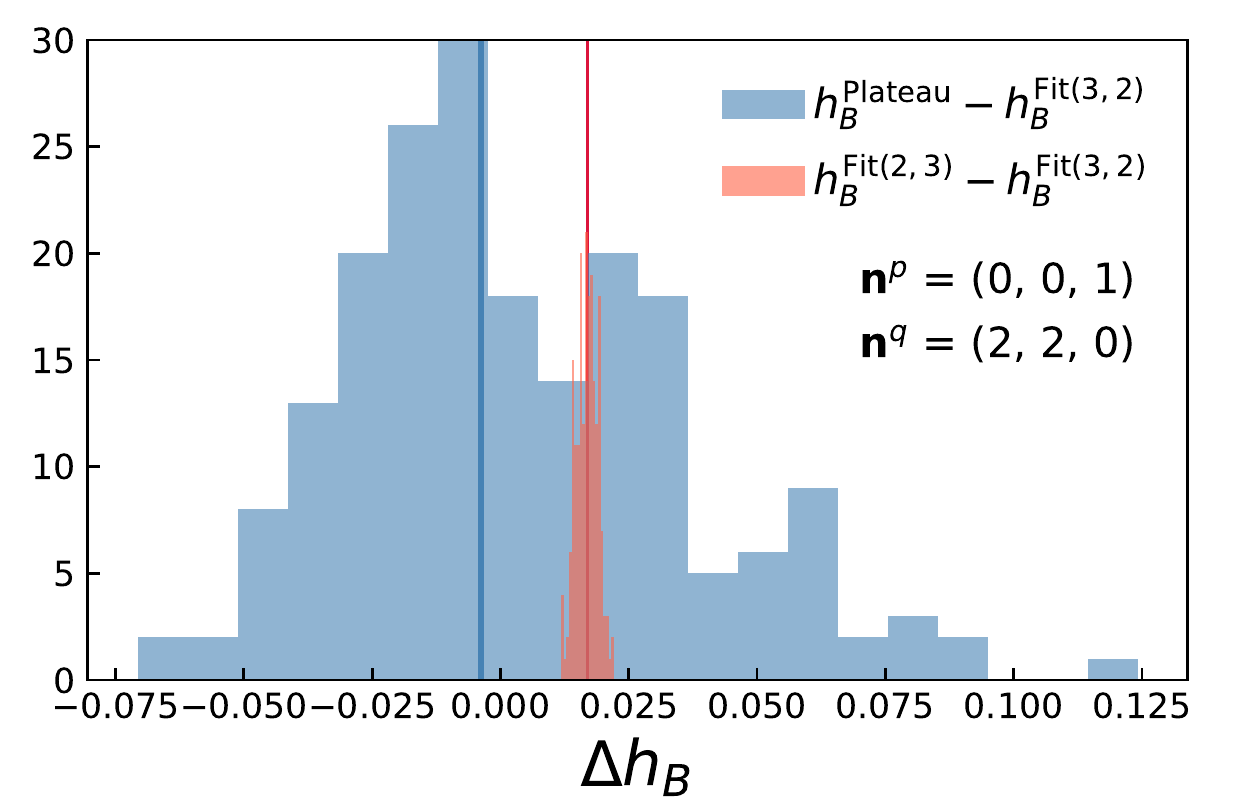}
\caption{The distributions of the bootstrap samples of $\Delta h_B$ between plateau fit and multi-state fit are shown for $\mathbf{n^q}=(0,0,0),~(1,0,0),~(1,1,0),~(2,0,0),~(2,1,0),~(2,2,0)$, where the vertical lines are the median values. \label{fig:fitdiff}}
\end{figure*}

It has been observed in \sec{c3pt} that the ratio $R^{fi}(\tau,t_s)$ of $t_s$ = 20a shows plateau around $t_s/2$ which is also consistent with the results from Fit(3,2) method, implying that the smallness of excited-state contribution in this region. Therefore it is reasonable to perform a one-state fit, namely plateau fit, to extract the bare matrix elements. We denote this method by Plateau($\tau_{\rm min},~\tau_{\rm max}$) which fit $R^{fi}(\tau,t_s=20a)$ of $\tau \in$ [$\tau_{\rm min},~\tau_{\rm max}$] to a constant.

The fit results from Plateau($\tau_{\rm min},~\tau_{\rm max}$) are shown in \fig{plateaufit} as the blue bands where the multi-state fit results are also shown for comparison. Clearly, the plateau fit shows good agreement with 3-state fit results. In \fig{fitdiff}, we show the distribution of difference between plateau fit and multi-state fit using bootstrap samples. In the main text, we have taken the difference between 2-state and 3-state fit as the systematic errors of excited-state contamination. It can be seen that such an estimate is larger than the difference between plateau fit and 3-state fit which should give a sufficiently conservative total error.

\begin{figure}
\includegraphics[width=0.45\textwidth]{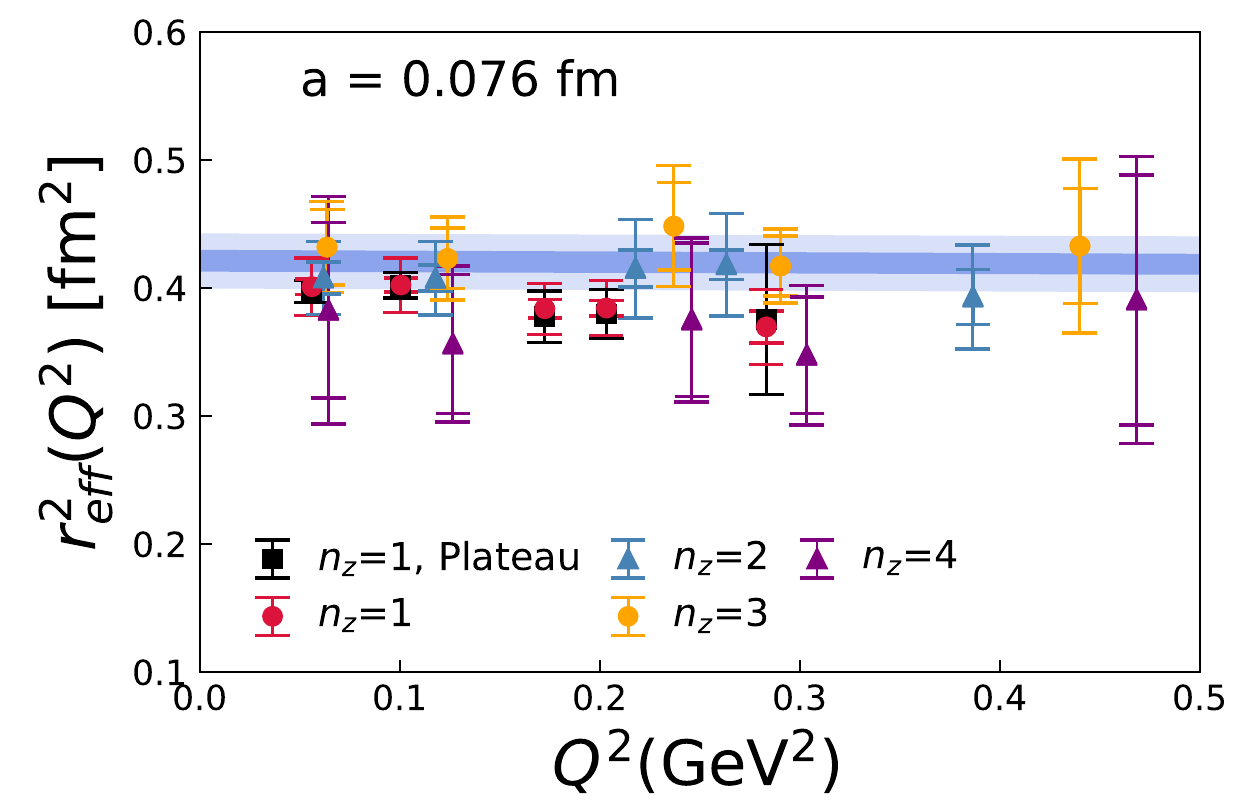}
\caption{Similar plot to \fig{Effradius} for $a=0.076$ fm ensemble including the results from plateau fit of $n_z$ = 1.\label{fig:EffradiusApp}}
\end{figure}

We also determined the pion
form factor from the plateau fits for $t_s=20$ The corresponding results in terms of the
effective radius are shown in \fig{EffradiusApp}.  Once again, consistent results between Plateau($\tau_{\rm min},~\tau_{\rm max}$) and Fit(3,2) can be observed. 

\section{Model dependence of radius extraction}\label{app:app3}
\begin{figure}
\includegraphics[width=0.45\textwidth]{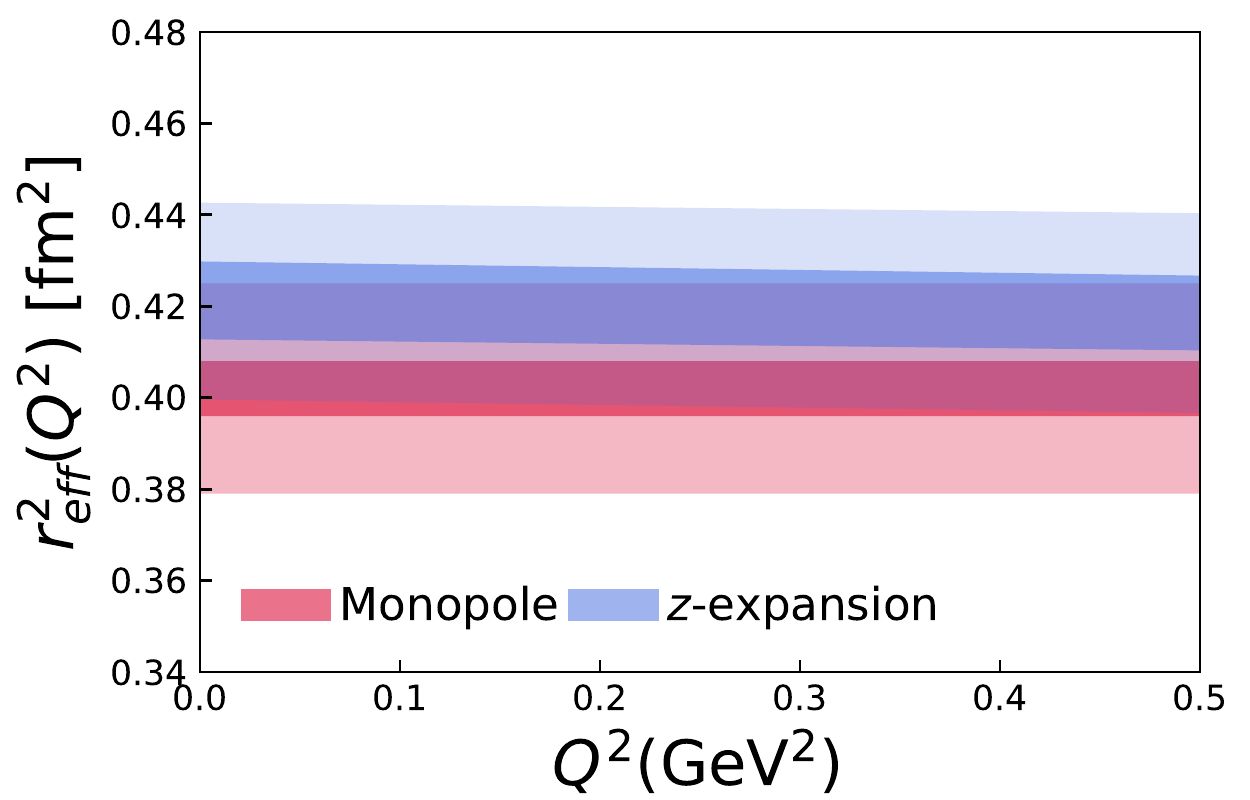}
\caption{The blue band is constructed by solving \Eq{FFradius} using $z$-expansion fit results as a function of $Q^2$, while the red band is a constant from monopole fit. The darker bands are the statistic errors from 3-state fit, while the lighter bands also include the systematic errors from the difference between 2-state and 3-state fit.\label{fig:effradiusdiff}}
\end{figure}

\begin{figure}
\includegraphics[width=0.45\textwidth]{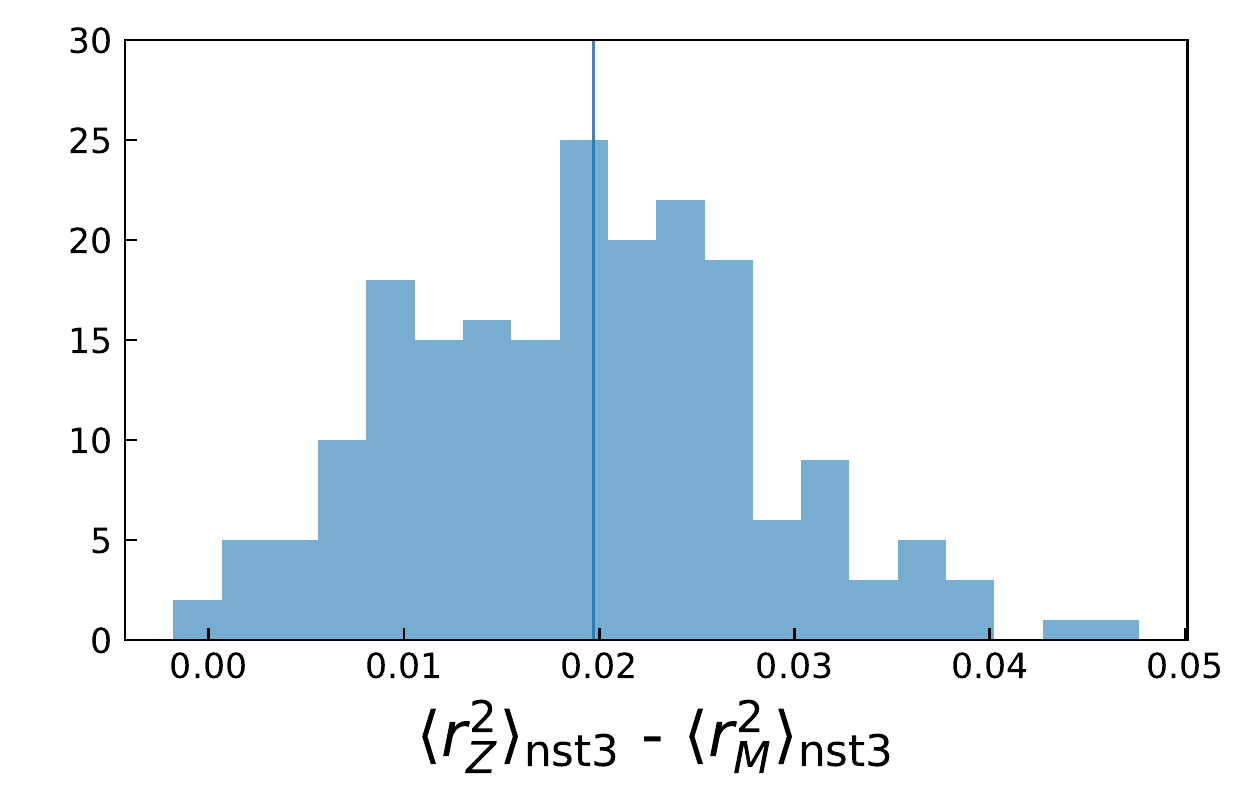}
\caption{Distribution of systematic errors $\langle r^2_{M} \rangle_{\rm nst3} - \langle r^2_{Z} \rangle_{\rm nst3}$ from bootstrap samples, where the N-state fit is denoted by nstN.\label{fig:radiusdiff}}
\end{figure}

In this work, we used $z$-expansion Ansatz to obtain the charge radius from the pion form factors shown in \tb{radius}. For comparison, in \tb{radiusM} we also show the radius obtained from monopole fit whose statistical error are often smaller, but this fit has larger systematic errors compared to $z$-expansion. Both fits produce good $\chi^2/df$. For the a = 0.076 fm ensemble, for example we get,
 $\chi^2/df=0.56$ for monopole fit, and 
 $\chi^2/df=0.51$ for $z$-expansion fit. Within the estimated errors the two fit forms give consistent results but only marginal. In \fig{effradiusdiff}, we show the effective radius (c.f. \Eq{Effradius}) calculated from the $z$-expansion fit (blue band) as well as monopole fit (red band). Clearly the $z$-expansion fit is more flexible so that the effective radius is a function of $Q^2$ rather than a constant. At $Q^2$ = 0 where the charge radius is defined, the result from $z$-expansion fit ($\langle r^2_{Z} \rangle$) is higher than monopole fit ($\langle r^2_{M} \rangle$). We show the distribution of  $\langle r^2_{M} \rangle_{\rm nst3} - \langle r^2_{Z} \rangle_{\rm nst3}$ from bootstrap samples in \fig{radiusdiff}, where the N-state fit is denoted by nstN. The central value of this distribution is 0.02 $\rm fm^2$.
 
 \begin{table}
\centering
\begin{tabular}{|c|c|c|c|c|c|c|}
\hline
\hline 
Data&$n_z$&$\langle r^2_{M} \rangle$ [$\rm{fm^2}$]\\
\hline 
a=0.076fm&[1,3]&0.402(6)(23)\\
\hline 
a=0.06fm&[0,3]&0.339(4)(18)\\
\hline 
a=0.04fm&[1,3]&0.313(5)(27)\\
\hline
\hline
\end{tabular}
\caption{The charge radius computed from monopole fit ($\langle r^2_{M} \rangle$). The first error is statistical, while the second error is systematic.
}\label{tb:radiusM}
\end{table}

%

\end{document}